\documentclass[ijoc]{informs4}

\usepackage{amsmath,amssymb,amsfonts,mathrsfs}
\usepackage{natbib}
\usepackage{bibunits}
\usepackage[bookmarks=true,hypertexnames=false,linkcolor=red,citecolor=blue,urlcolor=green,colorlinks=true,breaklinks]{hyperref}
\usepackage{float}
\usepackage{flafter}
\usepackage{booktabs}
\usepackage{threeparttable}
\usepackage{array}
\usepackage{multirow}
\usepackage{subfigure}
\usepackage{tikz}
\usepackage{xcolor}
\usepackage{colortbl}
\usepackage{listings}
\usepackage[T1,OT1]{fontenc} \usepackage{textcomp}
\usepackage{anyfontsize}
\usepackage{etoolbox}

\usetikzlibrary{arrows.meta,positioning,fit,calc,backgrounds,shapes.geometric}

\defaultbibliography{references.bib}
\defaultbibliographystyle{informs2014}

\newtheorem{prop}{Proposition}
\newcommand{\solver}{\textbf{PrecPack}}
\newcommand{\revisedbbr}{R-BBR23}
\newcommand{\enumsixteen}{EA16}
\newcommand{\enumseventeen}{EA17}
\newcommand{\stepref}[1]{Step~\textcolor{red}{#1}}

\hypersetup{pdftitle={\solver: An Efficient Open-Source Exact Solver for Bin Packing with Generalized Precedence Constraints},pdfauthor={Anonymous Authors}}

\definecolor{codeblue}{RGB}{42,82,145}
\definecolor{codegray}{RGB}{185,185,185}
\newcommand{\lightgraycell}{\cellcolor[rgb]{.949, .949, .949}}

\makeatletter
\renewcommand{\theARTICLEABSTRACT}{\HOOKb
  \vspace*{18pt}
  \begin{minipage}[t]{\textwidth}\parindent1em
    \ABSfont
    \noindent\theABSTRACT\endgraf
    \vskip5pt
    \theFUNDING
    \theKEYWORDS
    \theSUBJECTCLASS
    \theAREAOFREVIEW
    \theMSCCLASS
    \theORMSCLASS
    \if@BLINDREV\else\theHISTORY\fi
    \noindent\hrulefill
  \end{minipage}\vspace*{0pt}}
\makeatother

\lstdefinestyle{precpackcpp}{
  language=C++,
  basicstyle=\fontencoding{T1}\ttfamily\small,
  keywordstyle=\color{codeblue}\bfseries,
  commentstyle=\color{black!60}\itshape,
  stringstyle=\color{teal!65!black},
  frame=single,
  framesep=3pt,
  framerule=0.35pt,
  rulecolor=\color{codegray},
  backgroundcolor=\color{white},
  columns=fullflexible,
  keepspaces=true,
  tabsize=2,
  upquote=true,
  showstringspaces=false,
  breaklines=true,
  aboveskip=2pt,
  belowskip=2pt,
  xleftmargin=4pt,
  xrightmargin=4pt
}

\newcommand{\ManuscriptLineSpacing}{1.43}

\makeatletter
\newcommand{\ApplyManuscriptLineSpacing}[1]{\@tempdima=#1pt\relax
  \@tempdima=\ManuscriptLineSpacing\@tempdima
  \@tempdima=\dimexpr\@tempdima*2/3\relax
  \baselineskip=\@tempdima plus 1pt minus 1pt\relax
  \@bls=\@tempdima
}
\apptocmd{\AAnormalsizeXI}{\ApplyManuscriptLineSpacing{19}}{}{}
\apptocmd{\AAnormalsizeXII}{\ApplyManuscriptLineSpacing{20.7}}{}{}
\apptocmd{\AAsmallX}{\ApplyManuscriptLineSpacing{17.25}}{}{}
\apptocmd{\AAfootnotesizeIX}{\ApplyManuscriptLineSpacing{15.5}}{}{}
\makeatother

\OneAndAHalfSpacedXI

\def\bibfont{\fontsize{7.5pt}{10pt}\selectfont}
\def\bibsep{\smallskipamount}

\bibpunct[, ]{(}{)}{,}{a}{}{,}
\graphicspath{{Images/}}

\newcommand{\FloatTextGap}{12pt}
\newcommand{\FloatBetweenGap}{12pt}
\newcommand{\compactdisplays}{\setlength{\abovedisplayskip}{5pt}\setlength{\belowdisplayskip}{5pt}\setlength{\abovedisplayshortskip}{3pt}\setlength{\belowdisplayshortskip}{3pt}}
\appto{\normalsize}{\compactdisplays}
\appto{\small}{\compactdisplays}
\appto{\footnotesize}{\compactdisplays}

\TheoremsNumberedThrough
\EquationsNumberedThrough
\JOURNAL{INFORMS Journal on Computing}
\MANUSCRIPTNO{}

\newif\ifShowJournalHeader
\ShowJournalHeadertrue
\newcommand{\HiddenJournalHeaderLift}{57pt}
\let\PrecPackOriginalArticleTop\theARTICLETOP
\renewcommand{\theARTICLETOP}{\ifShowJournalHeader
    \PrecPackOriginalArticleTop
  \else
    \vspace*{-\HiddenJournalHeaderLift}\fi
}

\begin{document}
\begin{bibunit}

\RUNAUTHOR{Wang et al.}
\RUNTITLE{\solver: An Efficient Open-Source Exact Solver for BPP-GP}
\TITLE{\fontsize{18pt}{20pt}\selectfont \solver: An Efficient Open-Source Exact Solver for Bin Packing with Generalized Precedence Constraints}

\ARTICLEAUTHORS{
  \AUTHOR{Sunkanghong Wang$^{a,e}$, Zhengzhong Ricky You$^{b}$, Roberto Baldacci$^{c}$, Baichuan Mo$^{b}$, Hu Qin$^{d}$,\\ Lijun Wei$^{e,*}$, Zhou Xu$^{a}$}

  \vspace{0.5em}

  \AFF{$^a$Department of Logistics and Maritime Studies, Faculty of Business,
  The Hong Kong Polytechnic University, Hong Kong, China}

  \AFF{$^b$Department of Civil Engineering, Tsinghua University, Beijing 100084, China}

  \AFF{$^c$College of Science and Engineering, Hamad Bin Khalifa University, Doha, Qatar}

  \AFF{$^d$School of Management, Huazhong University of Science \& Technology, Wuhan, China}

  \AFF{$^e$Guangdong Provincial Key Laboratory of Computer Integrated Manufacturing, Guangdong University of Technology, Guangzhou, China}

  \AFF{$^*$Corresponding author}
}

\ABSTRACT{Efficient resource use in packing and assembly-line applications requires decisions that jointly account for capacity and precedence constraints. The strongly $\mathcal{NP}$-hard bin packing problem with generalized precedence constraints (BPP-GP) models such decisions by minimizing the number of ordered, capacitated bins required to pack weighted items, even when precedence requirements span multiple bins. Existing exact algorithms primarily focus on classical special cases, whereas general BPP-GP has been addressed only via compact integer models and heuristics, with no efficient open-source exact solver. We present \solver, a unified exact solver that extends branch-bound-and-remember (BBR) to arbitrary nonnegative precedence weights and naturally specializes to the classical cases. Generalized states capture restrictions that remain active across future bins, which are addressed through branching, dominance, and conflict-aware lower bounds. Root column generation uses fixed-point arithmetic to compute numerically valid dual bounds for pruning or to prove optimality. To support reuse and verification, we provide common programming and command-line interfaces, independent assignment checking, explicit termination statuses, and reproducible batch execution; the core procedures require no commercial software. In same-machine, single-threaded comparisons on classic assembly-line benchmarks, more instances are proven optimal, and average computing times are substantially reduced relative to leading source-available BBR implementations. Further comparisons with published benchmark results for bin packing with precedence constraints and BPP-GP also show that more instances were proved optimal and that reported average gaps were smaller on most benchmark sets. \solver\ is released under the MIT License at \href{https://github.com/Sunkanghong-Wang/PrecPack}{\texttt{https://github.com/Sunkanghong-Wang/PrecPack}}.}

\FUNDING{This work was supported by the Natural Science Foundation of China [Grants 72671084, 72271062, and 52575565], Science and Technology Projects in Guangzhou [Grant 2024A04J01019], and the Province Natural Science Fund of Guangdong [Grants 2025A1515110258, 2025A1515011297, and 2024A1515010246].}


\maketitle

\section{Introduction}\label{sec:introduction}

The classic bin packing problem (BPP) minimizes the number of identical capacitated bins used to pack weighted items and is strongly $\mathcal{NP}$-hard by reduction from 3-Partition \citep{garey1979computers}. The BPP with generalized precedence constraints (BPP-GP) inherits this hardness and additionally imposes generalized precedence constraints across ordered bins. Packing an item may make a successor ineligible for the same bin and, depending on the precedence weight, for several subsequent bins. Capacity and precedence decisions are therefore inseparable, and a partial packing over the bins already fixed must retain more information than the set of items already assigned.

\begin{figure}[htbp]
\centering
\subfigure[Generalized Precedence Graph]{\label{fig:example:a}\begin{tikzpicture}[
x=0.60cm,
y=0.58cm,
tasknode/.style={circle,draw,line width=0.55pt,minimum size=6.2mm,inner sep=0pt,font=\small},
weightlabel/.style={font=\fontsize{7.5pt}{8pt}\selectfont,inner sep=0pt},
precedencearc/.style={-{Latex[length=1.6mm,width=1.2mm]},line width=0.55pt},
arcweight/.style={font=\scriptsize,fill=white,inner sep=0.7pt}
]
\node[tasknode] (i1) at (0,8) {1};
\node[tasknode] (i6) at (3,8) {6};
\node[tasknode] (i10) at (6,8) {10};
\node[tasknode] (i16) at (12,8) {16};
\node[tasknode] (i2) at (0,6) {2};
\node[tasknode] (i7) at (3,6) {7};
\node[tasknode] (i11) at (6,6) {11};
\node[tasknode] (i13) at (9,6) {13};
\node[tasknode] (i17) at (12,6) {17};
\node[tasknode] (i3) at (0,4) {3};
\node[tasknode] (i18) at (12,4) {18};
\node[tasknode] (i4) at (0,2) {4};
\node[tasknode] (i8) at (3,2) {8};
\node[tasknode] (i12) at (6,2) {12};
\node[tasknode] (i14) at (9,2) {14};
\node[tasknode] (i20) at (12,2) {20};
\node[tasknode] (i5) at (0,0) {5};
\node[tasknode] (i9) at (3,0) {9};
\node[tasknode] (i15) at (9,0) {15};
\node[tasknode] (i19) at (12,0) {19};
\draw[precedencearc] (i1) -- node[arcweight] {1} (i6);
\draw[precedencearc] (i6) -- node[arcweight] {0} (i10);
\draw[precedencearc] (i10) -- node[arcweight,pos=0.48] {3} (i13);
\draw[precedencearc] (i2) -- node[arcweight] {0} (i7);
\draw[precedencearc] (i7) -- node[arcweight] {3} (i11);
\draw[precedencearc] (i11) -- node[arcweight] {1} (i13);
\draw[precedencearc] (i13) -- node[arcweight,pos=0.52] {1} (i16);
\draw[precedencearc] (i13) -- node[arcweight] {0} (i17);
\draw[precedencearc] (i13) -- node[arcweight,pos=0.52] {0} (i18);
\draw[precedencearc] (i4) -- node[arcweight] {0} (i8);
\draw[precedencearc] (i8) -- node[arcweight] {4} (i12);
\draw[precedencearc] (i12) -- node[arcweight] {1} (i14);
\draw[precedencearc] (i12) -- node[arcweight,pos=0.48] {0} (i15);
\draw[precedencearc] (i14) -- node[arcweight] {1} (i20);
\draw[precedencearc] (i5) -- node[arcweight] {1} (i9);
\draw[precedencearc] (i15) -- node[arcweight] {2} (i19);
\node[weightlabel,above=0.5mm of i1] {$w_1=8$};
\node[weightlabel,above=0.5mm of i2] {$w_2=6$};
\node[weightlabel,above=0.5mm of i3] {$w_3=8$};
\node[weightlabel,above=0.5mm of i4] {$w_4=4$};
\node[weightlabel,above=0.5mm of i5] {$w_5=9$};
\node[weightlabel,above=0.5mm of i6] {$w_6=7$};
\node[weightlabel,above=0.5mm of i7] {$w_7=3$};
\node[weightlabel,above=0.5mm of i8] {$w_8=14$};
\node[weightlabel,above=0.5mm of i9] {$w_9=7$};
\node[weightlabel,above=0.5mm of i10] {$w_{10}=2$};
\node[weightlabel,above=0.5mm of i11] {$w_{11}=7$};
\node[weightlabel,above=0.5mm of i12] {$w_{12}=4$};
\node[weightlabel,above=0.5mm of i13] {$w_{13}=2$};
\node[weightlabel,above=0.5mm of i14] {$w_{14}=9$};
\node[weightlabel,above=0.5mm of i15] {$w_{15}=7$};
\node[weightlabel,above=0.5mm of i16] {$w_{16}=11$};
\node[weightlabel,above=0.5mm of i17] {$w_{17}=5$};
\node[weightlabel,above=0.5mm of i18] {$w_{18}=12$};
\node[weightlabel,above=0.5mm of i19] {$w_{19}=13$};
\node[weightlabel,above=0.5mm of i20] {$w_{20}=4$};
\end{tikzpicture}}
\hfill
\subfigure[Optimal Seven-Bin Solution]{\label{fig:example:b}\begin{tikzpicture}[
x=0.205cm,
y=0.76cm,
binborder/.style={line width=0.55pt},
binlabel/.style={font=\small,inner sep=0pt},
residual/.style={fill=black!35}
]
\foreach \binindex/\binrow in {1/6,2/5,3/4,4/3,5/2,6/1,7/0} {
\node[binlabel,anchor=east,xshift=-2mm] at (0,{\binrow+0.34}) {\binindex};
}

\fill[residual] (27,6) rectangle (30,6.68);
\draw[binborder] (0,6) rectangle (30,6.68);
\foreach \x in {4,13,27} {\draw[binborder] (\x,6) -- (\x,6.68);}
\node[binlabel] at (2,6.34) {4};
\node[binlabel] at (8.5,6.34) {5};
\node[binlabel] at (20,6.34) {8};
\fill[residual] (24,5) rectangle (30,5.68);
\draw[binborder] (0,5) rectangle (30,5.68);
\foreach \x in {8,14,17,24} {\draw[binborder] (\x,5) -- (\x,5.68);}
\node[binlabel] at (4,5.34) {1};
\node[binlabel] at (11,5.34) {2};
\node[binlabel] at (15.5,5.34) {7};
\node[binlabel] at (20.5,5.34) {9};
\fill[residual] (17,4) rectangle (30,4.68);
\draw[binborder] (0,4) rectangle (30,4.68);
\foreach \x in {8,15,17} {\draw[binborder] (\x,4) -- (\x,4.68);}
\node[binlabel] at (4,4.34) {3};
\node[binlabel] at (11.5,4.34) {6};
\node[binlabel] at (16,4.34) {10};
\fill[residual] (0,3) rectangle (30,3.68);
\draw[binborder] (0,3) rectangle (30,3.68);
\fill[residual] (18,2) rectangle (30,2.68);
\draw[binborder] (0,2) rectangle (30,2.68);
\foreach \x in {7,11,18} {\draw[binborder] (\x,2) -- (\x,2.68);}
\node[binlabel] at (3.5,2.34) {11};
\node[binlabel] at (9,2.34) {12};
\node[binlabel] at (14.5,2.34) {15};
\fill[residual] (28,1) rectangle (30,1.68);
\draw[binborder] (0,1) rectangle (30,1.68);
\foreach \x in {2,11,16,28} {\draw[binborder] (\x,1) -- (\x,1.68);}
\node[binlabel] at (1,1.34) {13};
\node[binlabel] at (6.5,1.34) {14};
\node[binlabel] at (13.5,1.34) {17};
\node[binlabel] at (22,1.34) {18};
\fill[residual] (28,0) rectangle (30,0.68);
\draw[binborder] (0,0) rectangle (30,0.68);
\foreach \x in {11,24,28} {\draw[binborder] (\x,0) -- (\x,0.68);}
\node[binlabel] at (5.5,0.34) {16};
\node[binlabel] at (17.5,0.34) {19};
\node[binlabel] at (26,0.34) {20};
\foreach \x in {0,...,30} {\draw[line width=0.35pt] (\x,0) -- (\x,-0.18);}
\foreach \x in {0,11,24,28,30} {\draw[binborder] (\x,0) -- (\x,-0.22);}
\end{tikzpicture}}
\caption{A BPP-GP Instance with $n=20$ and $C=30$ and an Optimal Packing with an Empty Fourth Bin}
\label{fig:example}
\end{figure}
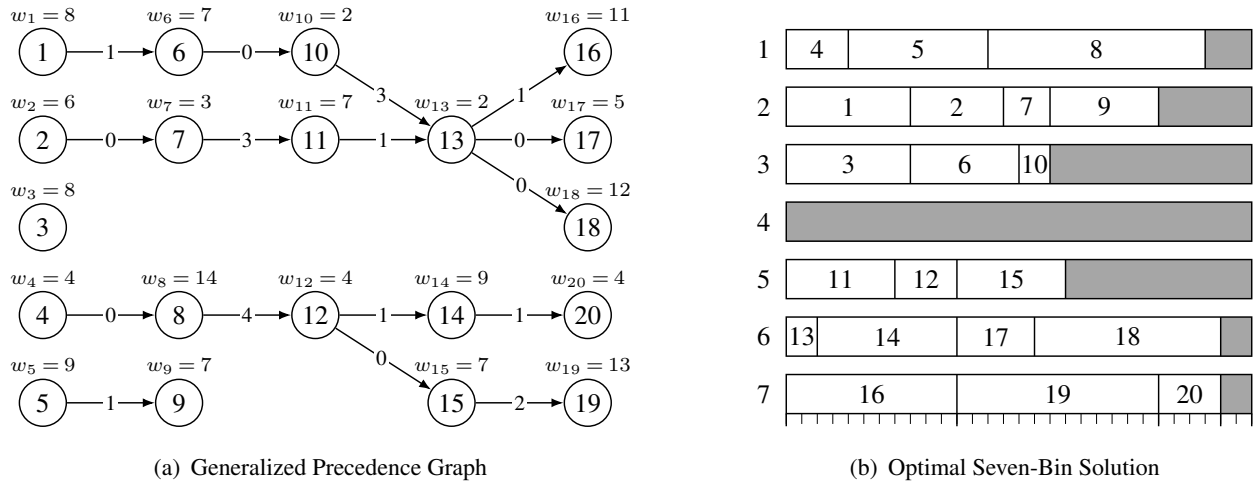

Formally, let $\mathcal{I}=\{1,\ldots,n\}$ be the item set, with $n\ge1$, and let $C$ be the positive integer bin capacity. Each item $i\in\mathcal{I}$ has a positive integer weight $w_i\le C$. Bins are available in an unlimited number and are indexed by positive integers. Generalized precedence is represented by the acyclic digraph $\mathcal{G}=(\mathcal{I},\mathcal{A})$, where $\mathcal{A}$ is the set of precedence arcs. Each arc $(i,j)\in\mathcal{A}$ carries a nonnegative integer precedence weight $t_{ij}$, the minimum required difference between the bin index of $j$ and that of $i$. A solution is an assignment of items to bins, represented by the vector $\mathbf{b}=(b_1,\ldots,b_n)\in\mathbb{Z}_{>0}^{n}$, where $b_i$ is the index of the bin to which item $i$ is assigned. Solution $\mathbf{b}$ is feasible if the total weight assigned to each bin does not exceed the capacity,
$\sum_{i\in\mathcal{I}:\,b_i=k} w_i\le C, \; \forall k\in\mathbb{Z}_{>0}$, and if it satisfies the generalized precedence constraints 
\begin{equation}
b_j-b_i\ge t_{ij},\qquad \forall (i,j)\in\mathcal{A}.
\label{eq:precedence-weight}
\end{equation}
The objective is to find a feasible solution that minimizes $\max_{i\in\mathcal{I}} b_i$, i.e., the index of the last bin. Since precedence weights larger than one may force some bins to remain empty, this index can exceed the number of nonempty bins, even in an optimal solution.
Figure~\ref{fig:example} shows an instance with 20 items and capacity $C=30$, along with an optimal packing using seven bins, where the fourth bin is empty. The numbers to the left of the bins indicate their indices, from 1 at the top to 7 at the bottom. Removing the empty bin and moving subsequent bins forward would violate precedence requirements. The possibility of intermediate empty bins distinguishes BPP-GP from BPP and complicates the adaptation of existing BPP techniques.

The BPP-GP generalizes two classical problems as special cases. When $t_{ij}=0$ for every arc, predecessors and successors may share a bin, yielding the simple assembly line balancing problem of type I (SALBP-I), where items are assembly tasks, weights are processing times, bins are stations, and capacity is the prescribed cycle time \citep{sewell2012branch}. When $t_{ij}=1$ for every arc, successors must occupy strictly later bins, yielding the bin packing problem with precedence constraints (BPP-P). Its applications include multiprocessor scheduling of unit-duration tasks with a single renewable resource \citep{garey1976resource}, electronics assembly with part-before-shield requirements \citep{tirpak2008developing}, and field-programmable gate array reconfiguration for image processing \citep{augustine2009strip}. BPP-GP further captures heterogeneous waiting requirements in production to preserve structural properties \citep{kramer2017batching} and in construction to let structures gain strength before further assembly \citep{chassiakos2005time}.

Let $H$ be an upper bound on the optimal BPP-GP solution value, e.g., the smallest possible index of the last bin, and let $\mathcal{B}=\{1,\ldots,H\}$ be the set of bin indices. The formulation represents a solution as an ordered sequence of bins. Binary variable $x_{ib}$ equals one if item $i$ is assigned to bin $b$, and binary variable $y_b$ equals one if the bin in position $b$ is used, including any required intermediate empty bin. A compact BPP-GP formulation proposed by \citet{kramer2017batching} is
\begin{subequations}\label{eq:compact}
\begin{alignat}{2}
\min\quad &\sum_{b\in\mathcal{B}}y_b, & \label{eq:compact-objective}\\
\text{s.t.}\quad &\sum_{b\in\mathcal{B}}x_{ib}=1, && \qquad \forall i\in\mathcal{I}, \label{eq:compact-assignment}\\
&\sum_{i\in\mathcal{I}}w_i x_{ib}\le Cy_b, && \qquad \forall b\in\mathcal{B}, \label{eq:compact-capacity}\\
&\sum_{b\in\mathcal{B}}b x_{jb}-\sum_{b\in\mathcal{B}}b x_{ib}\ge t_{ij}, && \qquad \forall (i,j)\in\mathcal{A}, \label{eq:compact-precedence}\\
&y_b\ge y_{b+1}, && \qquad \forall b\in\{1,\ldots,H-1\}, \label{eq:compact-prefix}\\
&x_{ib}\in\{0,1\}, && \qquad \forall i\in\mathcal{I},\ b\in\mathcal{B}, \label{eq:compact-x}\\
&y_b\in\{0,1\}, && \qquad \forall b\in\mathcal{B}. \label{eq:compact-y}
\end{alignat}
\end{subequations}

Constraints \eqref{eq:compact-assignment} assign each item exactly once, Constraints \eqref{eq:compact-capacity} enforce capacity limits, and Constraints \eqref{eq:compact-precedence} impose precedence-weight requirements. Constraints \eqref{eq:compact-prefix} account for each bin through the last occupied one, including intermediate empty bins required by positive precedence weights. Although the compact formulation can be solved by general-purpose integer programming solvers, this approach is generally limited to proving optimality for very small instances, as shown in Section \ref{sec:published-summary}. This limited scalability motivates the development of efficient, dedicated exact solvers for BPP-GP.

\subsection{Related Work and the Gap in Open-Source Exact Solvers}\label{sec:exact-review}

The literature on exact methods for SALBP-I and BPP-P provides the closest methodological background. For SALBP-I, exact algorithms have evolved from early laser and iterative-deepening searches through SALOME to cyclic best-first search (CBFS), branch-bound-and-remember (BBR), bounded dynamic programming (DP), and later improved variants. For BPP-P, research has progressed from bidirectional depth-first search (DFS) within branch-and-bound (BB) to CBFS-based BBR enumeration. In particular, both problems are also addressed by the branch-cut-and-price (BCP) method of \citet{letelier2022timelags}.

\begin{table}[t]
\centering
\caption{Representative Exact Algorithms and Solvers for SALBP-I and BPP-P}
\fontsize{8pt}{8pt}\selectfont
\renewcommand\arraystretch{1.2}
\tabcolsep=5pt
\resizebox{\textwidth}{!}{
\begin{tabular}{cllllc}
\toprule
Problem & Solver name & Reference & Search framework & Branching or item-subset generation & Public code \\
\midrule
\multirow[c]{13}{*}{SALBP-I} & FABLE88 & \citet{johnson1988optimally} & Laser search (DFS) & Item branching for the earliest bin & \\
& OptPack91 & \citet{nourie1991finding} & Laser search (DFS) & Item branching for the earliest bin & \\
& EUREKA92 & \citet{hoffmann1992eureka} & Iterative-deepening BB & {Item subsets in forward/reverse runs} & \\
& SALOME97 & \citet{scholl1997salome} & DFS with local lower-bound method & Bidirectional item-subset generation & \\
& SALOME99 & \citet{scholl1999balancing} & DFS with local lower-bound method & Bidirectional item-subset generation & \\
& BBR12 & \citet{sewell2012branch} & CBFS, then breadth-first BBR & {Item subsets in one chosen direction} & $\checkmark$ \\
& BBR14 & \citet{morrison2014application} & CBFS, then breadth-first BBR & {Item subsets in one chosen direction} & $\checkmark$ \\
& I-SALOME20 & \citet{li2020comparative} & {Modified CBFS-based BB} & Bidirectional item-subset generation & \\
& I-BDP20 & \citet{li2020comparative} & {Bounded DP with beam search} & {Item subsets in one chosen direction} & \\
& I-BBR20 & \citet{li2020comparative} & Modified CBFS-based BBR & {Item subsets in one chosen direction} & \\
& I-BBR20$'$ & \citet{li2020comparative} & Beam-search BBR & {Item subsets in one chosen direction} & \\
& BCP22 & \citet{letelier2022timelags} & BCP & Ryan--Foster item-pair branching & \\
& \revisedbbr & \citet{alvarezmiranda2023analysis} & CBFS, then breadth-first BBR & {Item subsets in one chosen direction} & $\checkmark$ \\
\midrule
\multirow[c]{4}{*}{BPP-P} & BB12 & \citet{dell2012bin} & Bidirectional DFS-based BB & {Bidirectional item-subset generation} & \\
& \enumsixteen & \citet{pereira2016procedures} & CBFS-based BBR & {Item subsets in one chosen direction} & \\
& \enumseventeen & \citet{kramer2017batching} & CBFS-based BBR & {Item subsets in one chosen direction} & \\
& BCP22 & \citet{letelier2022timelags} & BCP & Ryan--Foster item-pair branching & \\
\bottomrule
\end{tabular}}
\begin{tablenotes}
\fontsize{6.8pt}{7.5pt}\selectfont
\item \textit{Notes.} I-SALOME20, I-BDP20, I-BBR20, and I-BBR20$'$ denote ISALOME, IBDP, IBBR1, and IBBR2, respectively, in \citet{li2020comparative}. \enumseventeen\ denotes the rerun of \enumsixteen\ after implementation fixes, as reported by \citet{kramer2017batching}. BCP22 denotes the BCP-K variant of \citet{letelier2022timelags}, which is initialized with the best-known upper bounds obtained by \citet{kramer2017batching}.
\end{tablenotes}
\label{tab:review-summary}
\end{table}

Table~\ref{tab:review-summary} summarizes representative search frameworks and branching rules used in leading exact methods for SALBP-I and BPP-P; ordinary BPP and production-rate-maximization variants of SALBP-I are outside its scope. A checkmark denotes publicly downloadable application source code. This designation is distinct from open-source distribution under an explicit license. The BBR companion package includes the standard BBR12 and backtracking BBR14 implementations, while R-BBR23 is distributed via a public repository.
Dantzig--Wolfe lower bounds based on column generation have been developed for SALBP-I \citep{peeters2006linear,pereira2015empirical} and BPP-P \citep{pereira2016procedures}. The BCP method of \citet{letelier2022timelags} also uses a pattern-based relaxation with precedence-implied conflicts. However, none of these methods carefully address numerical issues arising from floating-point linear programming (LP) solvers, column generation may yield invalid dual bounds, potentially leading to incorrect pruning or an incorrect declaration of optimality, and may even enter an infinite iteration loop \citep{baldacci2024numerically,dasilva2025cutting}.

For BPP-GP, \citet{kramer2017batching} introduced the problem and proposed the compact bin-indexed formulation \eqref{eq:compact}, preprocessing procedures, adapted lower bounds, and the batching-move iterated local search (BM-ILS) heuristic. As discussed above, the compact model is exact but has limited computational scalability. BM-ILS can produce strong incumbents but is not an exact algorithm. \citet{letelier2022timelags} later developed an exact BCP method for special cases of bin packing with time lags. It can solve BPP-GP with weights in $\{0,1\}$, including SALBP-I and BPP-P, but cannot directly address general BPP-GP with arbitrary precedence weights.
Moreover, existing BBR methods for SALBP-I and BPP-P can be extended, but their state representations and dominance rules cannot be transferred unchanged. Under generalized precedence constraints, partial packings with the same assigned items can impose different restrictions on future bins. These restrictions must therefore be retained when comparing and remembering search states.

The methodological gap also concerns software availability. No explicitly licensed exact solver is available for arbitrary heterogeneous precedence weights. A research implementation is most useful when its mathematical state, termination statuses, independent validation procedure, and experimental protocol are inspectable. Related open-source work provides several examples. PyVRP and Alkaid-SDVRP combine high performance with documented reusable implementations \citep{wouda2024pyvrp,lin2026alkaid}; VRPSolverEasy exposes a sophisticated exact method through a compact interface \citep{errami2024vrpsolvereasy}; and RouteOpt combines exact performance with modular algorithm components and reproducible artifacts \citep{you2026routeopt}. Addressing this gap requires an exact BPP-GP solver that combines competitive performance with reusable interfaces, independently verifiable results, and reproducible experiments.

\subsection{Technical Foundations, Contributions, and Organization of This Paper}\label{sec:intro-contributions}

To address these gaps, we present \solver, a unified, efficient, open-source exact solver for BPP-GP that also specializes naturally to SALBP-I and BPP-P when all precedence weights are zero and one, respectively. It combines numerous classical and advanced techniques from the literature as follows:

\begin{itemize}

\item \textbf{\textit{Branching, search, and dominance.}} Branches are defined by maximal feasible item subsets \citep{dell2012bin,pereira2016procedures}. The search is organized using CBFS and remembering, while redundant branches are eliminated by the Jackson and no-successors rules \citep{sewell2012branch,morrison2014application}. States are ordered by the one-machine priority rule, and additional dominance tests are based on the one-item-superset and directed subgraph-isomorphism rules \citep{pereira2016procedures}.

\item \textbf{\textit{Preprocessing and primal heuristics.}} Precedence strengthening and item lifting tighten the precedence and capacity constraints, respectively \citep{kramer2017batching}. Direction selection aims to reduce branching near the root, and structured reductions simplify the instance before search \citep{pereira2016procedures}. First-fit and best-fit heuristics construct initial feasible packings, while modified Hoffmann heuristics \citep{hoffmann1963assembly,sewell2012branch,morrison2014application} and bounded DP \citep{pereira2016procedures} seek better incumbents.

\item \textbf{\textit{Lower bounds.}} To support search pruning and optimality proofs, we use precedence-path and window bounds \citep{kramer2017batching}, one-machine scheduling bounds \citep{pereira2016procedures}, the BPP-P $L_4$ network bound \citep{dell2012bin} computed with the \textsc{Dinic} algorithm \citep{dinic1970algorithm}, dual-feasible-function (DFF) bounds \citep{fekete2001new,boschetti2003two,haouari2005fast,crainic2007new}, and the ordinary-bin-packing lower-bound routine (BINLB) \citep{sewell2012branch}.

\item \textbf{\textit{Root relaxation.}} Pattern-based set covering provides a relaxation for strengthening the root lower bound \citep{gilmore1961linear,lubbecke2005selected}. Conflict-knapsack pricing identifies improving patterns that satisfy capacity and pairwise incompatibility constraints \citep{hifi2007reduction,pereira2016procedures,letelier2022timelags}, while a diversification strategy seeks multiple distinct improving patterns per iteration \citep{dasilva2025cutting}. Fixed-point arithmetic ensures that the root lower bound used for pruning or optimality decisions is numerically valid \citep{baldacci2024numerically}.

\end{itemize}

Building on these foundations, this study makes the following four unique contributions:

\begin{enumerate}
\renewcommand{\labelenumi}{(\arabic{enumi})}

\item \textbf{\textit{Generalized exact search.}} To the best of our knowledge, this study provides the first BBR-based exact method for BPP-GP with arbitrary nonnegative precedence weights and the last-bin-index objective. Its state representation retains precedence restrictions across future bins. Forced empty-bin transitions and generalized state and item dominance enable exact search under these restrictions. State-dependent precedence information is incorporated into the residual path, one-machine, and closure bounds, while pairwise conflicts strengthen BINLB and capacity tightening. Bounds are evaluated in order of computational cost, and auxiliary searches are limited to control the additional effort without compromising exactness.

\item \textbf{\textit{Computational performance.}} In same-machine, single-thread tests with the same time limit, \solver\ proves optimality for 523 of the 525 SALBP-I benchmark instances with 100 items, versus 456 or 457 for each of three source-available BBR implementations. The average computing time over all instances is 1.79 seconds for the proposed solver, compared with 50.39 to 55.18 seconds for the three implementations. Further comparisons with published state-of-the-art results for BPP-P and BPP-GP show that \solver\ proves optimality for more instances and achieves smaller average gaps on most benchmark sets.

\item \textbf{\textit{Reusable and reliable implementation.}} The C++ implementation provides a common application programming interface (API) and command-line interface (CLI), with separately testable algorithmic components, independent assignment checking, and regression tests. Pruning and optimality decisions rely on integer or discrete calculations. Fixed-point arithmetic ensures the numerical validity of the root lower bound computed from column-generation dual values. Explicit statuses distinguish optimality from resource-limited termination, and the core procedures require no commercial software.

\item \textbf{\textit{Reproducible experiments.}} Documented instance and solution formats, resumable batch scripts, and structured results allow users to repeat the comparisons and evaluate algorithmic extensions through a common experimental workflow. Algorithmic settings and applicability rules are specified in advance for all three problem classes, without benchmark-specific algorithm selection.

\end{enumerate}

The remainder of this paper is organized as follows. Section~\ref{sec:algorithm} presents the solution procedure, emphasizing the extended BBR search. Section~\ref{sec:software} describes the software design, implementation, and reproducibility. Section~\ref{sec:experiments} reports the computational experiments, and Section~\ref{sec:conclusion} gives conclusions and future research directions. The Online Supplement provides additional algorithmic details, proofs, and experimental settings.

\section{The \solver\ Solver}\label{sec:algorithm}

Let $\mathrm{LB}$ and $\mathrm{UB}$ denote the best valid lower bound and the value of the best incumbent found, respectively. The design principle of \solver\ is to apply inexpensive procedures first to strengthen bounds. These procedures may prove optimality directly or reduce the work required by the subsequent complete BBR search. The overall solution procedure is summarized as follows.

\begin{itemize}
\item \textbf{Step 1 (Input Validation, Preprocessing, and Initial Bounds):} Read the instance, check the consistency of its item, capacity, and precedence data, and apply the preprocessing procedures. Set $\mathrm{LB}$ to the maximum of several inexpensive initial lower bounds and initialize $\mathrm{UB}$ with the bin count of the best feasible packing obtained by constructive heuristics (see \S \ref{sec:initialization}).

\item \textbf{Step 2 (Preliminary Search and Incumbent Improvement):} Apply additional primal heuristics, followed by a limited preliminary BBR search and, when applicable, the bounded DP heuristic, to seek a smaller $\mathrm{UB}$. If the preliminary BBR search proves optimality, set $\mathrm{LB}=\mathrm{UB}$ (see \S \ref{sec:incumbent-improvement}).

\item \textbf{Step 3 (One-Fewer-Bin Test and Direction Choice):} Compute the earliest and latest admissible bins for each item under the candidate limit $\mathrm{UB}-1$. If any resulting interval is empty, no solution using at most $\mathrm{UB}-1$ bins exists, so set $\mathrm{LB}=\mathrm{UB}$. Otherwise, use these intervals to choose whether to retain the current precedence graph or reverse all its arcs, aiming to reduce branching in the subsequent search (see \S \ref{sec:one-fewer-bin-test}).

\item \textbf{Step 4 (Root-Bound Strengthening by Column Generation):} Apply column generation to a special relaxation without explicit bin positions and increase $\mathrm{LB}$ if a stronger lower bound is obtained (see \S \ref{sec:root-strengthening}).

\item \textbf{Step 5 (Complete BBR Search):} Conduct a complete BBR search to improve $\mathrm{UB}$ and strengthen $\mathrm{LB}$, stopping when optimality is proved or the configured time or memory limit is reached (see \S \ref{sec:bbr}).

\item \textbf{Step 6 (Reconstruction, Validation, and Reporting):} Restore the input item numbering, verify the feasibility of the best solution found, and report the solution, final bounds, and whether optimality was proved or the run was stopped because the time or memory limit was reached (see \S \ref{sec:reconstruction-validation}).
\end{itemize}

Whenever $\mathrm{LB}=\mathrm{UB}$ during the solution process, optimality is proved, so the remaining steps are skipped and execution proceeds directly to \stepref{6}. The following subsections describe these steps in detail.

\subsection{Input Validation, Preprocessing, and Initial Bounds}\label{sec:initialization}

Before preprocessing, we verify that $n=|\mathcal{I}|\ge 1$, $C\in\mathbb{Z}_{>0}$, $w_i\in\{1,\ldots,C\}$ for all $i\in\mathcal{I}$, $\mathcal{A}\subseteq\{(i,j)\in\mathcal{I}\times\mathcal{I}:i\ne j\}$, $t_{ij}\in\mathbb{Z}_{\ge 0}$ for all $(i,j)\in\mathcal{A}$, and that $\mathcal{G}=(\mathcal{I},\mathcal{A})$ is acyclic. Multiple input arcs with the same ordered endpoints are replaced by a single arc with the maximum precedence weight. Instances that fail any of these conditions are rejected before preprocessing.

After the input passes these checks, each preprocessing round begins with the precedence-strengthening procedure of \citet{kramer2017batching}, which uses the current precedence and capacity constraints to derive stronger precedence relations between items, thereby strengthening the graph $\mathcal{G}$. Then, the bin capacity $C$ is additionally tightened using a conflict-aware maximum-fill calculation introduced in this study. Specifically, for any $\widehat{\mathcal{I}}\subseteq\mathcal{I}$, let $W(\widehat{\mathcal{I}})=\sum_{i\in\widehat{\mathcal{I}}}w_i$, and call two items compatible if the strengthened graph contains no positive-weight precedence relation between them in either direction. The tightened capacity is
\begin{equation}
C'=\max\left\{W(\widehat{\mathcal{I}}):\widehat{\mathcal{I}}\subseteq\mathcal{I},\ W(\widehat{\mathcal{I}})\le C,\ \widehat{\mathcal{I}}\text{ is pairwise compatible}\right\}
\label{eq:preprocessing-capacity}
\end{equation}
and is computed by a standard BB algorithm that branches on whether to include each item and uses subset-sum upper bounds \citep{hifi2007reduction}. Because the items assigned to any feasible bin form a pairwise-compatible subset considered in Eq.~\eqref{eq:preprocessing-capacity}, the total weight of every feasible bin is at most $C'$; thus, replacing $C$ with $C'$ preserves feasibility and optimality. Based on the tightened capacity, the item-lifting procedure of \citet{kramer2017batching} and the direction rule of \citet{pereira2016procedures} are then applied. Next, the items are renumbered in topological order, with ties broken by nonincreasing weight and then by the nonincreasing number of immediate successors. Thus, a complete preprocessing round applies precedence strengthening, capacity tightening, item lifting, direction selection, and item renumbering in that order. Another round starts from the updated instance if and only if at least one of these procedures changes the instance, and all five procedures are executed again. The direction rule in this loop establishes a working orientation for the remaining initialization procedures; the one-fewer-bin test in Section \ref{sec:one-fewer-bin-test} may later reverse this orientation before the structured reductions and exact-search phases. The structured reductions of \citet{pereira2016procedures} are outside this loop and, when every input precedence arc has weight one, are applied after the final orientation is selected.

After preprocessing, the initial lower bound $\mathrm{LB}$ is calculated as the maximum of the continuous BPP bound \citep{dell2012bin}, the precedence-path bounds \citep{kramer2017batching}, and the window-DFF bounds \citep{kramer2017batching}.
For an input designated as BPP-P, the longest-path maximum-flow bound $L_4$ of \citet{dell2012bin} is additionally computed. Next, an initial upper bound is constructed using a fit-based heuristic that evaluates the preprocessed topological item order, a stable nonincreasing-weight order that retains the topological order for ties, and 20 shuffled orders. At each placement, only unassigned items whose predecessors have already been assigned are considered. For each order, one packing is constructed using first-fit and another using best-fit, with the latter permitting only feasibility-preserving exchanges. The better packing is retained, and the best solution found across all evaluated orders initializes $\mathrm{UB}$.
The strengthened graph is used to compute the initial bounds and in the fit-based, modified Hoffmann, and bounded DP primal heuristics, as well as in the one-fewer-bin test. In contrast, the limited preliminary BBR search, root relaxation, and complete BBR search use the original precedence relations. Excluding the additional strengthened relations in these searches prevents the maximum precedence weight from increasing and, consequently, the size and memory requirements of each search state. Our preliminary experiments also indicated better average performance with this choice.

\subsection{Preliminary Search and Incumbent Improvement}\label{sec:incumbent-improvement}

If an optimality gap remains, the modified Hoffmann heuristic, adapted from \citet{sewell2012branch}, first evaluates a set of scoring functions that combine item weight and successor information.
For each scoring function, a packing is constructed one bin at a time by enumerating a capped set of capacity- and precedence-feasible item subsets for each bin and selecting the highest-scoring subset; the best resulting packing is retained. For SALBP-I, the item-subset enumeration follows \citet{sewell2012branch} and successively uses limits of 50, 250, 500, and 1000 subsets per bin when $n\le 200$ and a single limit of 50 when $n>200$. To extend this heuristic from SALBP-I to BPP-GP, the feasibility checks used during subset enumeration are modified to enforce arbitrary nonnegative precedence weights. An item may share a bin with a predecessor only when the connecting arc has weight zero; otherwise, their bin indices must satisfy the prescribed minimum difference.
A limited BBR search is then run for at most 0.05 seconds to solve easy instances. For BPP-P and BPP-GP, a bounded DP heuristic based on \citet{pereira2016procedures} subsequently advances one bin per stage and retains a ranked subset of the generated successor states. To accommodate generalized precedence constraints, its state representation additionally records restrictions that remain active across future bins, distinguishing partial packings with the same assigned items but different future restrictions. The transition rules enforce the required differences between bin indices and create an empty bin when no nonempty feasible subset of items can be assigned. It keeps at most 1000 states per stage, 50 promising transitions per state, and 1000 generated item subsets per state. Let $T$ denote the global time limit. The modified Hoffmann heuristic, the limited BBR search, and the bounded DP heuristic share an initialization budget of $\min\{3,0.1T\}$ seconds, measured from the start of preprocessing.

\subsection{One-Fewer-Bin Test and Direction Choice}\label{sec:one-fewer-bin-test}

To efficiently determine whether the incumbent is optimal, the one-fewer-bin test targets $K=\mathrm{UB}-1$ bins. Following the earliest/latest-bin construction of \citet{pereira2016procedures}, it computes a valid lower bound $\underline{b}_i$ on the bin index of each item $i$ from terms based on precedence paths and the capacity needed by transitive predecessors, as well as a valid lower bound $f_i$ on the number of bins required after the bin containing $i$ from the corresponding successor terms. The resulting valid upper bound on its bin index is $\overline{b}_i=K-f_i$. If $\underline{b}_i>\overline{b}_i$ for some item $i\in\mathcal{I}$, no solution can use at most $K$ bins, and the incumbent is therefore optimal. Otherwise, the direction is selected by comparing two products. The forward product multiplies the counts $|\{i\in\mathcal{I}:\underline{b}_i\le b\}|$ over the first at most five bins $b$, whereas the reverse product multiplies $|\{i\in\mathcal{I}:\overline{b}_i\ge b\}|$ over the last at most five bins. The working orientation is reversed only if the reverse product is smaller. For any positive integer horizon $H$, reversing every precedence arc and mapping $b_i$ to $H+1-b_i$ defines a feasibility-preserving bijection between packings within bins $1,\ldots,H$ under the two orientations. Thus, feasibility across all horizons, and hence the optimal bin count, remains unchanged. This choice determines only the orientation used in the subsequent reductions, root-bound strengthening, and complete BBR search.

\subsection{Root Lower-Bound Strengthening by Column Generation}\label{sec:root-strengthening}

To further strengthen $\mathrm{LB}$ before the complete BBR search, we solve a set-covering relaxation without explicit bin indices by column generation. The relaxation and the subsequent BBR search use the instance obtained after preprocessing and any applicable structured reductions. If structured reductions have fixed bins outside this instance, their number is subtracted from $\mathrm{LB}$ and $\mathrm{UB}$ for these calculations and restored when reporting bounds and solutions; a nonempty remaining instance has a lower bound of at least one.

Let $\bar t_{ij}$ be the maximum sum of precedence weights over all directed paths from item $i$ to a distinct item $j$, with $\bar t_{ij}=0$ if no such path exists, and let $\mathcal{I}_{ij}$ contain the endpoints and every item on a directed path from $i$ to $j$. The relaxation uses the conflict graph $\mathcal{G}^{\mathrm{c}}=(\mathcal{I},\mathcal{E}^{\mathrm{c}})$, where, for distinct items $i,j$, $\{i,j\}\in\mathcal{E}^{\mathrm{c}}$ if $\bar t_{ij}>0$, $\bar t_{ji}>0$, or there exists $(u,v)\in\{(i,j),(j,i)\}$ such that item $v$ is reachable from item $u$ and $W(\mathcal{I}_{uv})>C$. A pattern is an item subset that respects capacity and contains no edge in $\mathcal{E}^{\mathrm{c}}$; let $\mathscr{P}$ be the set of all such subsets. For each $p\in\mathscr{P}$, let $a_{ip}$ equal one if pattern $p$ contains item $i$ and zero otherwise, let $\lambda_p$ be its fractional use, and let $h\in\mathbb{R}$ be the bin-count variable. The position-free LP relaxation is
\begin{subequations}\label{eq:pf-m}
\begin{alignat}{2}
\min\quad &h, & \label{eq:pf-m-objective}\\
\text{s.t.}\quad &\sum_{p\in\mathscr{P}}a_{ip}\lambda_p\ge 1, && \quad \forall i\in\mathcal{I}, \label{eq:pf-m-cover}\\
&\sum_{p\in\mathscr{P}}\lambda_p\le h, && \label{eq:pf-m-count}\\
&\mathrm{LB}\le h\le\mathrm{UB},\quad \lambda_p\ge 0, && \quad \forall p\in\mathscr{P}. \label{eq:pf-m-domain}
\end{alignat}
\end{subequations}
Relaxation \eqref{eq:pf-m} follows the modeling approach of \citet{letelier2022timelags}, whose patterns incorporate pairwise incompatibilities implied by weighted precedence paths. Their BCP method applies column generation at the nodes of a BB tree and adds cycle-elimination inequalities. In contrast, we apply column generation only at the root to strengthen $\mathrm{LB}$, without subsequently branching on this model. The cycle-elimination inequalities are also omitted, so pricing remains a conflict-knapsack problem.
Relative to their set-partitioning formulation, we use the covering inequalities \eqref{eq:pf-m-cover}, which restrict the associated dual multipliers to nonnegative values. We also minimize $h$ rather than the total number of nonempty patterns used. Constraint \eqref{eq:pf-m-count} allows $h$ to exceed this total, accommodating the intermediate empty bins counted by the BPP-GP objective without explicitly determining their positions.
Every feasible pattern belongs to $\mathscr{P}$: a positive-weight precedence path prevents its endpoints from sharing a bin, while placing related items $u$ and $v$ together forces all items in $\mathcal{I}_{uv}$ into that bin and therefore requires $W(\mathcal{I}_{uv})\le C$.
Every feasible packing using at most $\mathrm{UB}$ bins induces a feasible solution to Model \eqref{eq:pf-m}. Thus, the relaxation optimum provides a valid lower bound.

Because the set $\mathscr{P}$ may be exponentially large, Model \eqref{eq:pf-m} is solved by column generation. The restricted position-free relaxation is initialized with one pattern for each nonempty bin in the incumbent solution; intermediate empty bins produce no pattern because $h$ represents the total number of bins independently of the fractional use of nonempty patterns. These initial patterns cover every item and therefore yield a feasible initial restricted relaxation. Column generation then iteratively solves the restricted relaxation and augments it with negative-reduced-cost patterns returned by the pricing problem. Specifically, let $\pi_i\ge0$ denote the profit of item $i$ in the pricing problem, equal to the dual multiplier of Constraint \eqref{eq:pf-m-cover}, and let $\mu\ge0$ denote the dual multiplier of Constraint \eqref{eq:pf-m-count}, written as $h-\sum_{p\in\mathscr{P}}\lambda_p\ge0$. The reduced cost of a pattern $p$ is $\mu-\sum_{i\in\mathcal{I}}\pi_i a_{ip}$. The bounds on $h$ affect only the dual condition associated with $h$ and therefore do not enter the reduced cost of a pattern. Because $\mu$ is identical for all patterns, minimizing reduced cost is equivalent to maximizing pattern profit. For each item $i\in\mathcal{I}$, let $\xi_i$ be one if item $i$ is selected for the generated pattern and zero otherwise. The resulting conflict-knapsack pricing problem is as follows.
\begin{subequations}\label{eq:root-pricing}
\begin{alignat}{2}
V^\star(\boldsymbol{\pi})=\max\quad &\sum_{i\in\mathcal{I}}\pi_i \xi_i, & \label{eq:root-pricing-objective}\\
\text{s.t.}\quad &\sum_{i\in\mathcal{I}}w_i \xi_i\le C, && \label{eq:root-pricing-capacity}\\
&\xi_i+\xi_j\le 1, && \forall \{i,j\}\in\mathcal{E}^{\mathrm{c}}, \label{eq:root-pricing-conflict}\\
&\xi_i\in\{0,1\}, && \forall i\in\mathcal{I}. \label{eq:root-pricing-domain}
\end{alignat}
\end{subequations}
A negative-reduced-cost pattern exists only if $V^\star(\boldsymbol{\pi})>\mu$. The problem \eqref{eq:root-pricing} is solved by the DP-guided BB method of \citet{hifi2007reduction}. Items with positive profit are numbered $1,\ldots,r$, placing items with no incident conflict first and items involved in a conflict last. Let $\pi_{(j)}$ and $w_{(j)}$ denote the profit and weight of the $j$th item in this order. BB processes this order backward, so undecided items always form a prefix, and conflict-free items are processed last. Ignoring conflicts, let $\operatorname{DP}(j,c)$ be the maximum profit obtainable from the first $j$ items in this order with capacity $c$:
\begin{equation}
\operatorname{DP}(j,c)=
\begin{cases}
\operatorname{DP}(j-1,c), & c<w_{(j)},\\
\max\left\{\operatorname{DP}(j-1,c),\ \pi_{(j)}+\operatorname{DP}(j-1,c-w_{(j)})\right\}, & c\ge{}w_{(j)},
\end{cases}
\label{eq:pricing-dp}
\end{equation}
with $\operatorname{DP}(0,c)=0$ for $c=0,\ldots,C$; the recurrence is evaluated for every $j=1,\ldots,r$ and $c=0,\ldots,C$.

At a BB node with current profit $v$, residual capacity $c$, and the first $j$ items in this order still undecided, $v+\operatorname{DP}(j,c)$ is a valid upper bound even when some of these items conflict with selected items. When seeking to improve columns, a node is pruned if this bound does not exceed the larger of the best admissible pricing profit found and $\mu$. Once all items involved in a conflict have been decided, the remaining items have no conflicts with each other or with selected items, so the DP value gives the exact best completion and closes the node. A heuristic pricing procedure uses the diversification strategy of \citet{dasilva2025cutting} to generate a limited set of negative-reduced-cost patterns.

To prevent floating-point error from affecting the lower bound, the nonnegative dual values are scaled and rounded toward zero to obtain integer item profits, following the fixed-point principles of \citet{baldacci2024numerically}. For lower-bound calculation, a separate pricing call maximizes the scaled profit over the full pattern set using integer arithmetic. In this integer pricing call, the pruning threshold is the best feasible profit found, not the scaled column-generation threshold. The resulting lower bound is calculated from the scaled profits and the proven pricing optimum, rather than by rounding the objective value of the restricted relaxation; the integer calculation is given in Section \ref{app:root-integer-bound} of the Online Supplement.
Note that the column-generation procedure is valid for every instance but, based on preliminary experiments, is invoked only when $|\mathcal{I}|\le 100$. Its total budget is $\min\{0.2,\max\{0.05,0.0015|\mathcal{I}|\},0.05T_{\mathrm{rem}}\}$ seconds, where $T_{\mathrm{rem}}$ is the remaining global computing time; the procedure is skipped if this budget is below 0.05 seconds.

\subsection{BBR Search}\label{sec:bbr}

The core search procedure in \solver\ extends the BBR methods developed for SALBP-I \citep{sewell2012branch,morrison2014application} and the exact-search techniques developed for BPP-P \citep{pereira2016procedures} to accommodate generalized precedence constraints in BPP-GP. The search repeatedly selects a search state, assigns a feasible subset of items to the next bin, applies valid bounds and dominance rules, and stores previously explored states to avoid redundant work. The following steps summarize the procedure.

\begin{itemize}
  \item \textbf{Step 5.1 (Initialization):} Create the root state (see \S \ref{sec:state-representation}), initialize the exact-state table and the depth-indexed priority queues used by CBFS, and place the root state in the queue for depth 0 (see \S \ref{sec:search-memory}).

  \item \textbf{Step 5.2 (State Selection):} Select the next open state from the priority queues using CBFS (see \S \ref{sec:search-memory}). If no open state remains, the search is exhausted, and the incumbent is proved optimal.

  \item \textbf{Step 5.3 (Branch Generation):} Enumerate the maximal feasible item subsets for the next bin, ensuring no further unassigned items can be added without violating feasibility (see \S \ref{sec:maximal-load}). If some items remain unassigned and no nonempty feasible subset exists, generate the unique forced-empty-bin transition. During the enumeration process, apply item-replacement dominance rules (see \S \ref{sec:separation-dominance}) to eliminate dominated branches before constructing child states, thus reducing the number of states the search evaluates.

  \item \textbf{Step 5.4 (Child Evaluation and Insertion):} Construct a child state for each retained branch. If the child assigns every item, update the incumbent if its bin count is smaller than $\mathrm{UB}$. Otherwise, evaluate the inexpensive capacity and DFF bounds, consult the exact-state table (see \S \ref{sec:search-memory}), apply the dominance tests (see \S \ref{sec:separation-dominance}), and then evaluate the more expensive residual lower bounds (see \S \ref{sec:search-bounds}). Prune the child as soon as any test succeeds. If the child is not pruned, store it and add it to the queue for its depth. If the same state was previously stored at a greater depth, update that state and reinsert it at the smaller depth instead.

  \item \textbf{Step 5.5 (Continuation and Termination):} Continue processing the retained branches unless the time or memory limit is reached. After processing all branches of the current state, return to \stepref{5.2}. If the search is interrupted by reaching either limit, preserve the current incumbent and valid lower bound (see \S \ref{sec:search-initialization}). Record which limit caused termination and proceed to \stepref{6}.
\end{itemize}

The remainder of this subsection defines the BBR state for generalized precedence constraints and presents the branching, remembering, dominance, and lower-bound procedures used in these steps. It then describes search initialization and termination under resource limits.

\subsubsection{State Representation for Generalized Precedence Constraints.}\label{sec:state-representation}

At a bin boundary, the assigned-item set alone is sufficient to describe the remaining precedence restrictions in SALBP-I and BPP-P. However, in general BPP-GP, an assigned predecessor can keep a successor unavailable for several future bins, so the remaining separation distances must also be recorded. Consider a search state immediately before opening bin $d+1$, where bins $1,\ldots,d$ have been fixed and $d\in\mathbb{Z}_{\ge0}$. Let $\mathcal{I}^{\mathrm{A}}\subseteq\mathcal{I}$ be the assigned-item set and $\mathcal{I}^{\mathrm{U}}=\mathcal{I}\setminus\mathcal{I}^{\mathrm{A}}$ the unassigned-item set. Let $\tau=\max\{0,\max_{(i,j)\in\mathcal{A}}t_{ij}-1\}$ be the largest number of future bins over which a newly imposed precedence restriction can remain active, with the inner maximum equal to zero when $\mathcal{A}=\varnothing$. For each $r=1,\ldots,\tau$, define $\mathcal{D}_r=\{j\in\mathcal{I}^{\mathrm{U}}:\exists i\in\mathcal{I}^{\mathrm{A}},\ (i,j)\in\mathcal{A},\ b_i+t_{ij}>d+r\}$. These are the items excluded from all of the next $r$ bins by a direct arc from an already assigned predecessor. When $\tau\ge1$, the sets satisfy $\mathcal{D}_1\supseteq\mathcal{D}_2\supseteq\cdots\supseteq\mathcal{D}_{\tau}$; for notational convenience, let $\mathcal{D}_r=\varnothing$ for every $r>\tau$. Let $\mathscr{D}=(\mathcal{D}_1,\ldots,\mathcal{D}_{\tau})$ be the separation tuple.

The BBR state is then $S=(\mathcal{I}^{\mathrm{A}},\mathscr{D})$, serving as the key used by the exact-state table. When all precedence weights are zero or one, $\tau=0$ and $\mathscr{D}$ is empty, reducing the BBR state to the assigned-item set $\mathcal{I}^{\mathrm{A}}$. Let $\mathcal{L}\subseteq\mathcal{I}^{\mathrm{U}}$ denote the subset of items assigned to bin $d+1$, hereafter called a load, and let $\mathcal{I}^{\mathrm{A}\prime}=\mathcal{I}^{\mathrm{A}}\cup\mathcal{L}$. Feasibility requires $\sum_{i\in\mathcal{L}}w_i\le C$, and for every $j\in\mathcal{L}$, $j\notin\mathcal{D}_1$, every predecessor of $j$ connected by an arc of positive weight belongs to $\mathcal{I}^{\mathrm{A}}$, and every predecessor linked by an arc of weight zero belongs to $\mathcal{I}^{\mathrm{A}}\cup\mathcal{L}$. Thus, the endpoints of a precedence arc of weight zero may share a load, while every precedence constraint with positive weight is respected. The child separation tuple $\mathscr{D}'=(\mathcal{D}'_1,\ldots,\mathcal{D}'_{\tau})$ is
\begin{equation}
\mathcal{D}'_r=\left(\mathcal{D}_{r+1}\cup\left\{j\in\mathcal{I}:\exists i\in\mathcal{L},\ (i,j)\in\mathcal{A},\ t_{ij}>r\right\}\right)\setminus\mathcal{I}^{\mathrm{A}\prime},\qquad \forall r=1,\ldots,\tau,
\label{eq:separation-tuple-update}
\end{equation}
where $\mathcal{D}_{r+1}$ advances each existing restriction by one bin. The second set in \eqref{eq:separation-tuple-update} adds the restrictions created by the current load, while the final set difference excludes items that have already been assigned. Let $P=(\mathcal{L}_1,\ldots,\mathcal{L}_d)$ be a feasible partial packing: its loads are disjoint, meet capacity and precedence constraints among assigned items, and include all predecessors of assigned items. Each load may be empty while there are unassigned items; a complete packing ends at its last nonempty bin. Hence, a load corresponds to a single bin, whereas $P$ denotes the first $d$ fixed bins. Let $S_P$ denote the BBR state induced by $P$, and $\mathscr{C}(P)$ be the set of finite sequences of future loads that, starting with bin $d+1$, assign every remaining item exactly once, satisfying all capacity and precedence constraints, and ending when the last remaining item is assigned. If no items remain, $\mathscr{C}(P)$ contains only the empty sequence. Proposition~\ref{prop:state-sufficiency} shows that partial packings with the same BBR state admit the same feasible future-load sequences. This justifies remembering states without including the complete packing history in the state representation.

\begin{prop}[Sufficiency of the BBR State for BPP-GP]\label{prop:state-sufficiency}
For any two feasible partial packings $P$ and $\widetilde{P}$ of the same BPP-GP instance, $S_P=S_{\widetilde{P}}$ implies $\mathscr{C}(P)=\mathscr{C}(\widetilde{P})$.
\par\noindent\textit{Proof.} Given in Section \ref{app:proof-state} of the Online Supplement. \hfill\Halmos
\end{prop}

\subsubsection{Maximal-Load Branching and Forced Empty Bins.}\label{sec:maximal-load}

Consider a BBR state $S$ with at least one unassigned item. Let $\mathscr{F}(S)$ be the set of all nonempty loads feasible for the next bin, and let $\operatorname{Max}_{\subseteq}\mathscr{F}(S)$ denote the loads in $\mathscr{F}(S)$ that are maximal under set inclusion. The branching set is
\begin{equation}
\mathscr{L}(S)=
\begin{cases}
\operatorname{Max}_{\subseteq}\mathscr{F}(S), & \mathscr{F}(S)\neq\varnothing,\\
\{\varnothing\}, & \mathscr{F}(S)=\varnothing.
\end{cases}
\label{eq:branching-load-set}
\end{equation}
Precedence weights greater than one introduce a transition that does not arise in classical SALBP-I and BPP-P searches. If $\mathscr{F}(S)=\varnothing$, the next bin must be empty; the separation tuple is updated according to Eq.~\eqref{eq:separation-tuple-update}, and the bin is counted in the objective. If $\mathscr{F}(S)\neq\varnothing$, the empty load is excluded.

The principle of branching on maximal feasible loads follows \citet{dell2012bin} and \citet{pereira2016procedures}. We extend its feasibility conditions to account for the separation tuple and introduce a forced-empty-bin transition when no nonempty load is feasible. Proposition~\ref{prop:maximal-load} shows that these branches retain at least one minimum-length feasible future-load sequence, thereby justifying the restriction to maximal loads and forced empty bins. The procedure for enumerating maximal feasible loads, including within-bin eligibility updates, is detailed in Section~\ref{app:load-enumeration} of the Online Supplement.

\begin{prop}[Completeness of Generalized Branching]\label{prop:maximal-load}
For any BBR state $S$ generated by the search, let $\mathscr{C}(S)$ be the common set of feasible future-load sequences for partial packings that induce $S$, as established by Proposition \ref{prop:state-sufficiency}. If $\mathcal{I}^{\mathrm{U}}\neq\varnothing$, some minimum-length sequence in $\mathscr{C}(S)$ begins with a load in $\mathscr{L}(S)$.
\par\noindent\textit{Proof.} Given in Section \ref{app:proof-maximal-load} of the Online Supplement. \hfill\Halmos
\end{prop}

\subsubsection{CBFS and Exact-State Remembering.}\label{sec:search-memory}

CBFS selects the next open state. Open states are partitioned by depth, and the nonempty depth queues are visited cyclically. Within each depth, BPP-GP and its BPP-P special case use the ordering rule of \citet{pereira2016procedures}. For each $i\in\mathcal{I}^{\mathrm{U}}$, let $\eta_i$ be the maximum total precedence weight on a directed path starting at $i$ in the subgraph induced by $\mathcal{I}^{\mathrm{U}}$, and order the unassigned items as $u_1,\ldots,u_{|\mathcal{I}^{\mathrm{U}}|}$ so that $\eta_{u_1}\ge\cdots\ge\eta_{u_{|\mathcal{I}^{\mathrm{U}}|}}$. States with smaller values of $\max_{1\le k\le|\mathcal{I}^{\mathrm{U}}|}\{\sum_{\ell=1}^{k}w_{u_\ell}+C\eta_{u_k}\}$ are selected first, with ties first resolved in favor of the smaller value of $\min_{i\in\mathcal{I}^{\mathrm{U}}}\eta_i$.
In particular, for SALBP-I at depth $d>0$, we retain the BBR12 priority $(dC-W(\mathcal{I}^{\mathrm{A}}))/d-0.02|\mathcal{I}^{\mathrm{U}}|$, including the coefficient 0.02 used by \citet{sewell2012branch}. Our implementation multiplies this priority by $50d$ to obtain the integer value $50(dC-W(\mathcal{I}^{\mathrm{A}}))-d|\mathcal{I}^{\mathrm{U}}|$. Since states are compared within the same depth, this positive scaling preserves their order. Depth zero contains only the root.
Any remaining ties are resolved by the smaller state lower bound and then by state-generation order. These rules affect only the order of state expansion, not the generated state space or exactness.
For each BBR state $S$, exact-state remembering records the smallest depth at which $S$ has been reached. When $S$ is generated at depth $d$, it is discarded if the same state was previously reached at a depth $d'\le d$, because Proposition \ref{prop:state-sufficiency} gives both occurrences the same feasible future-load sequences. If the previous depth satisfies $d'>d$, the remembered depth is reduced, the depth-dependent lower bound is recalculated, and the state is returned to the queue at depth $d$.

\subsubsection{Dominance Rules.}\label{sec:separation-dominance}\label{sec:special-dominance}

Beyond exact-state remembering, \solver\ applies two dominance rules between BBR states and several between feasible item subsets for the next bin to reduce the search space. Consider $S=(\mathcal{I}^{\mathrm{A}},\mathscr{D})$ as an arbitrary reachable BBR state at depth $d$. A state is dominated by $S$ if every future-load sequence from it can be substituted by one from $S$ without increasing the objective value, allowing it to be discarded without losing an optimal solution.
The first rule, stated in Proposition~\ref{prop:tuple-dominance}, extends memory-based dominance \citep{sewell2012branch,pereira2016procedures} to states with identical assigned items but potentially different separation tuples. Under this rule, a state can be discarded if another state is reached at no greater depth and imposes no stronger restrictions on future bins.

\begin{prop}[Separation-Tuple Dominance]\label{prop:tuple-dominance}
Let $S'=(\mathcal{I}^{\mathrm{A}},\mathscr{D}')$, with $\mathscr{D}'=(\mathcal{D}'_1,\ldots,\mathcal{D}'_{\tau})$, be another reachable state of the same working instance at depth $d'$. If $d\le d'$ and $\mathcal{D}_r\subseteq\mathcal{D}'_r$ for every $r=1,\ldots,\tau$, then $S$ dominates $S'$.
\par\noindent\textit{Proof.} Given in Section \ref{app:proof-tuple} of the Online Supplement. \hfill\Halmos
\end{prop}

The second rule, stated in Proposition~\ref{prop:superset-dominance}, extends the maximum-load rule of \citet{pereira2016procedures} for SALBP-I and BPP-P. In these special cases, a state reached at no greater depth that has assigned the same items plus one dominates the state with fewer assigned items. For BPP-GP, this extension additionally requires the separation tuple of the dominating state to be componentwise contained in that of the dominated state.

\begin{prop}[Generalized Maximum-Load Dominance]\label{prop:superset-dominance}
For an item $i\in\mathcal{I}\setminus\mathcal{I}^{\mathrm{A}}$, let $S^{+}=(\mathcal{I}^{\mathrm{A}}\cup\{i\},\mathscr{D}^{+})$, with $\mathscr{D}^{+}=(\mathcal{D}^{+}_1,\ldots,\mathcal{D}^{+}_{\tau})$, be another reachable state of the same working instance at depth $d^{+}$. If $d^{+}\le d$ and $\mathcal{D}^{+}_r\subseteq\mathcal{D}_r$ for every $r=1,\ldots,\tau$, then $S^{+}$ dominates $S$.
\par\noindent\textit{Proof.} Given in Section \ref{app:proof-superset} of the Online Supplement. \hfill\Halmos
\end{prop}

The remaining rules compare feasible item subsets for the next bin. At the same state, one load dominates another if every feasible future-load sequence starting with the latter can be replaced by one starting with the former using no more bins. When all precedence weights are zero, the extended Jackson replacement and no-successors rules are applied \citep{sewell2012branch,morrison2014application}. When all precedence weights are one, transitive-successor containment and \citet{pereira2016procedures}'s directed subgraph-isomorphism dominance are used. The latter seeks an injective mapping from the selected item's successor subgraph to the replacement item's successor subgraph, preserving every arc and mapped nonroot weight while permitting a heavier replacement root. For heterogeneous precedence weights, unweighted successor containment is inadequate since two arcs to the same successor may impose different minimum separations. The labeled-successor item-dominance rule, stated in Proposition~\ref{prop:labeled-item-dominance}, prefers assigning an item with no smaller weight and no weaker successor requirements earlier, provided that the exchange is feasible; the item left for a later bin then has no larger weight and no stronger successor requirements. The item-replacement comparison is extended using the direct-arc label $\theta_{is}=t_{is}$ if $(i,s)\in\mathcal{A}$, and $\theta_{is}=-1$ otherwise. The condition $\theta_{is}\ge\theta_{js}$ for every $s\in\mathcal{I}$ requires each direct-successor restriction of item $j$ to be matched by one from item $i$ with at least the same precedence weight.
When applying this rule, equal weights and identical outgoing labels are resolved in favor of the smaller item index. In this case, $i$ may replace $j$ only if $i<j$. This convention prevents equivalent choices from eliminating one another.

\begin{prop}[Labeled-Successor Item Dominance]\label{prop:labeled-item-dominance}
Let $\mathcal{L}$ be a feasible maximal next load at state $S$, let $j\in\mathcal{L}$, and let $i\in\mathcal{I}\setminus(\mathcal{I}^{\mathrm{A}}\cup\mathcal{L})$. Define $\mathcal{L}'=(\mathcal{L}\setminus\{j\})\cup\{i\}$. If neither item is reachable from the other in $\mathcal{G}$, $w_i\ge w_j$, $\theta_{is}\ge\theta_{js}$ for every $s\in\mathcal{I}$, $\mathcal{L}'$ is feasible at $S$, and $i<j$ whenever $w_i=w_j$ and $\theta_{is}=\theta_{js}$ for every $s\in\mathcal{I}$, then there exists a feasible maximal next load $\widehat{\mathcal{L}}\supseteq\mathcal{L}'$ that dominates $\mathcal{L}$. Consequently, when replacements are applied only under these conditions, every discarded load is dominated by a retained load, and the branch beginning with $\mathcal{L}$ may be discarded.
\par\noindent\textit{Proof.} Given in Section \ref{app:proof-labeled-item} of the Online Supplement. \hfill\Halmos
\end{prop}

\subsubsection{Residual Lower Bounds.}\label{sec:search-bounds}

Residual (local) lower bounds prune states that cannot improve the incumbent. For a state $S$ at depth $d$, with $\mathcal{I}^{\mathrm{U}}\neq\varnothing$, define $z(S,d)=d+\min_{\boldsymbol{\mathcal{L}}\in\mathscr{C}(S)}|\boldsymbol{\mathcal{L}}|$ as its minimum achievable objective value. Each of the following bounds includes the $d$ fixed bins and is a valid lower bound on $z(S,d)$. The capacity and DFF bounds are examined first, followed by exact-state remembering and state dominance tests. For states that remain after these assessments, precedence-path, one-machine, closure, and bounded BINLB calculations are considered in that order. A state is discarded as soon as one of these bounds reaches $\mathrm{UB}$, without evaluating the remaining bounds.

Relaxing the precedence constraints for unassigned items yields the classic capacity and DFF bounds for BPP \citep{fekete2001new,boschetti2003two,haouari2005fast,crainic2007new}. A function $\phi:[0,1]\to[0,1]$ is dual feasible if $\sum_{k=1}^{m}\phi(x_k)\le1$ holds for every finite multiset $\{x_1,\ldots,x_m\}$ of positive real numbers with $\sum_{k=1}^{m}x_k\le1$. Let $\mathcal{T}_{\mathrm{DFF}}$ denote the retained set of DFFs.

For each $\phi\in\mathcal{T}_{\mathrm{DFF}}$, these bounds are
\begin{equation}
\mathrm{LB}_{\mathrm{cap}}=d+\left\lceil\frac{W(\mathcal{I}^{\mathrm{U}})}{C}\right\rceil,\qquad \mathrm{LB}_{\phi}=d+\left\lceil\sum_{i\in\mathcal{I}^{\mathrm{U}}}\phi(w_i/C)\right\rceil.
\label{eq:cheap-residual-bounds}
\end{equation}
At most seven DFFs are retained from a base candidate set. When all precedence weights are zero or all are one, up to eight additional DFFs are selected from an expanded candidate set; otherwise, only the base selection is used. Details are provided in Section \ref{app:dff-selection} of the Online Supplement.

To account for precedence, define $\rho_i(S)=\max(\{r\in\{1,\ldots,\tau\}:i\in\mathcal{D}_r\}\cup\{0\})$ for each unassigned item $i$. This quantity counts the consecutive bins, beginning with $d+1$, in which $i$ cannot be placed because of direct arcs from already assigned predecessors. Let $e_i$ be a valid lower bound on the offset of the bin assigned to $i$ relative to the next bin $d+1$. The value $\eta_i$ retains its meaning from Section \ref{sec:search-memory} as the maximum total precedence weight on a path from $i$ through unassigned successors. For every $i\in\mathcal{I}^{\mathrm{U}}$, the earliest offset and tail are calculated in topological and reverse-topological order, respectively, by
\begin{subequations}\label{eq:residual-offsets}
\begin{align}
e_i&=\max\left(\{\rho_i(S)\}\cup\{e_j+t_{ji}:(j,i)\in\mathcal{A},\ j\in\mathcal{I}^{\mathrm{U}}\}\right),\label{eq:release-offset}\\
\eta_i&=\max\left(\{0\}\cup\{t_{ij}+\eta_j:(i,j)\in\mathcal{A},\ j\in\mathcal{I}^{\mathrm{U}}\}\right).\label{eq:tail-offset}
\end{align}
\end{subequations}
Thus, the precedence-path bound is $\mathrm{LB}_{\mathrm{path}}=d+\max_{i\in\mathcal{I}^{\mathrm{U}}}(e_i+\eta_i+1)$. The BPP-P head and tail constructions of \citet{dell2012bin} and \citet{pereira2016procedures} use unit precedence weights, whereas \citet{kramer2017batching} define weighted heads and tails at the root of BPP-GP. Equations \eqref{eq:residual-offsets} extend this weighted construction to BPP-GP through $\rho_i(S)$, which accounts for restrictions imposed by already assigned items.

The same $e_i$ and $\eta_i$ serve in place of the root head and tail quantities in the forward and reverse one-machine bounds and the predecessor/successor closure bound \citep{dell2012bin,pereira2016procedures,kramer2017batching}. The one-machine bounds combine the number of bins required by the total weight of a subset with the requirements imposed by the earliest bin positions or remaining precedence paths of its items. Their state-dependent formulations and validity arguments are given in Section \ref{app:state-dependent-bounds} of the Online Supplement. In SALBP-I, all precedence weights and residual offsets are zero. Thus, $\mathrm{LB}_{\mathrm{path}}=d+1$ is dominated by the capacity bound, and both one-machine bounds equal the capacity bound; therefore, these three calculations are omitted. The closure bound is also evaluated for SALBP-I because it can still be stronger: the predecessors and successors of an item may each require several bins, and the corresponding bin intervals overlap only at the bin containing that item.

The final stage combines ordinary BINLB \citep{sewell2012branch} with a conflict-aware extension to test whether the unassigned items can fit into at most $\mathrm{UB}-d-1$ bins under relaxed precedence constraints. Proving infeasibility is sufficient to prune $S$ without determining the exact optimum of the relaxation. Ordinary BINLB removes all precedence constraints, resulting in a minimum bin count represented by $z_{\mathrm{BP}}(\mathcal{I}^{\mathrm{U}})$. The extension restricts the conflict graph $\mathcal{G}^{\mathrm{c}}$ from Section \ref{sec:root-strengthening} to $\mathcal{I}^{\mathrm{U}}$, leading to a bin-packing-with-conflicts relaxation whose optimum is denoted as $z_{\mathrm{BPPC}}(\mathcal{I}^{\mathrm{U}})$. This relaxation maintains capacity restrictions and pairwise incompatibilities while ignoring bin order and the placement of any necessary empty bins.

This stage begins by checking inexpensive capacity and DFF bounds, together with a clique bound when conflicts are present; each item in a clique requires a different bin. Unlike the earlier checks on the full BBR state, these bounds also apply to the smaller remaining subsets produced during the auxiliary searches. If the initial checks do not prune $S$, ordinary BINLB is applied. The exact conflict-aware search is considered only when the ordinary phase completes without pruning $S$. Because bin order is irrelevant in this relaxation, one selected item must be included in the next bin. Maximal capacity-feasible subsets containing that item and no conflicting pair are then enumerated, thereby removing bin symmetry.

If the auxiliary search proves that the unassigned items cannot fit into at most $k$ bins, the resulting lower bound is $d+k+1$. Otherwise, if an auxiliary optimum is obtained, the corresponding bound is $d+z_{\mathrm{BP}}(\mathcal{I}^{\mathrm{U}})$ or $d+z_{\mathrm{BPPC}}(\mathcal{I}^{\mathrm{U}})$. If an auxiliary search is interrupted, only bounds already established are retained, including a previously computed ordinary optimum; an unfinished test contributes no new infeasibility conclusion. The auxiliary searches use fixed effort limits, specified in Section \ref{app:conflict-binlb} of the Online Supplement.

\subsubsection{Search Initialization and Resource Limits.}\label{sec:search-initialization}

To reduce initialization time, we perform a brief BBR search before constructing the expanded DFF set and other expensive search components. The search uses the base DFF set, with generalized item dominance and BINLB disabled, and has a time budget of 0.001 seconds. It is attempted only when expanded DFF generation is enabled and applicable, and more than 0.002 seconds remain in the current BBR budget. The procedure is shared by the 0.05-second BBR search in Section \ref{sec:incumbent-improvement} and the complete BBR search; the short search consumes part of the corresponding budget and receives no additional time.

If the short search proves optimality, its result is returned immediately. Otherwise, if time remains in the enclosing BBR budget, only its validated incumbent is passed to a new search initialized at the root with the configured bounds and dominance rules. The stored states are not reused.
If the complete BBR search is interrupted by the global time or memory limit, the current incumbent and a valid lower bound are returned, together with the reason for termination.
A state remains open until its expansion is completed or its remaining branches are pruned, even after it has been removed from its priority queue. Thus, an interrupted expansion leaves the parent open to represent its ungenerated branches. Let $\mathscr{O}$ be the set of unfinished states, including any partially expanded parent, and let $\mathrm{LB}_0$ be the valid lower bound entering the BBR search. For each $S\in\mathscr{O}$, let $\ell(S)$ denote its retained lower bound on the total number of bins, including those already fixed, as described in Section~\ref{sec:search-bounds}. When $\mathscr{O}\neq\varnothing$, the reported lower bound is $\min\{\mathrm{UB},\max\{\mathrm{LB}_0,\min_{S\in\mathscr{O}}\ell(S)\}\}$. If search exhaustion is established, $\mathrm{LB}=\mathrm{UB}$; otherwise, if no unfinished state is available, $\mathrm{LB}_0$ is retained.
Section \ref{sec:resource-control} describes memory accounting.

\subsection{Reconstruction, Validation, and Reporting}\label{sec:reconstruction-validation}

Upon termination of the solution procedure, the best packing is reconstructed for the original instance. Any item renumbering or bin order reversal is undone, and items removed or preassigned by structured reductions are reinstated. Contributions from fixed bins excluded from root and BBR computations are added back into their bounds. The packing is verified using the original item weights, bin capacity, and precedence constraints to ensure that every item is assigned exactly once, that no bin exceeds capacity, and that all precedence constraints are met. The objective value is recalculated as the index of the largest occupied bin. Only packings that satisfy these conditions are returned, accompanied by the final lower and upper bounds. The status report indicates whether optimality was proved or if termination occurred due to time or memory limits.

\section{Software Design, Implementation, and Reproducibility}\label{sec:software}

The \solver\ described above is implemented in a common C++ core, available through a stand-alone executable and a library interface. Users can solve new instances, embed the solver in larger applications, and extend its algorithmic components while reusing the tools for instance handling and solution verification.

\subsection{Architecture and Public Interface}\label{sec:architecture-overview}

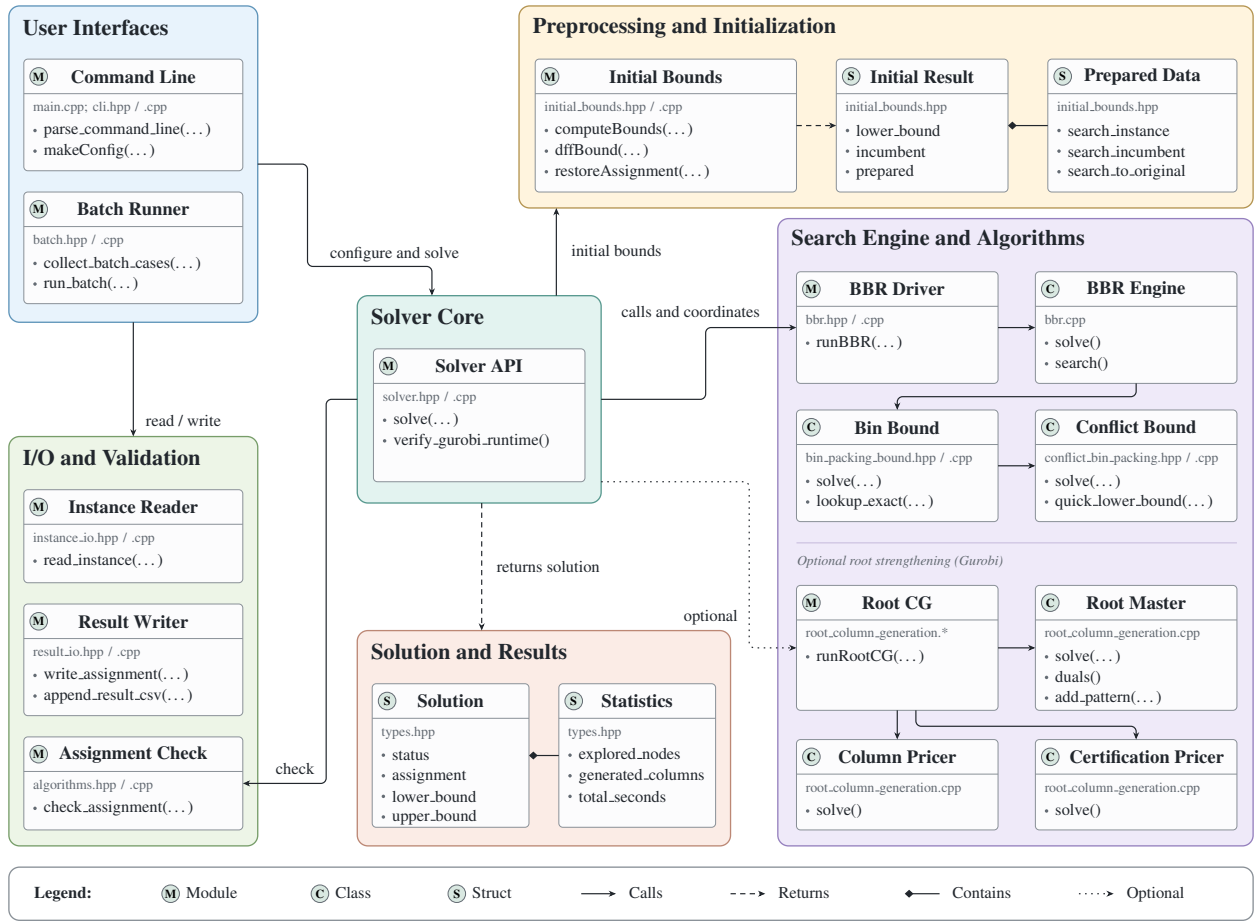
\begin{figure}[t]
\centering
\begingroup
\linespread{1}\selectfont

\def\SABodySize{6.2}
\def\SAClassSize{7.0}
\def\SAFileSize{5.2}
\def\SAGroupSize{8.4}
\def\SALabelSize{6.2}
\def\SAMarkerSize{4.6}
\def\SAMarkerDiameter{6.6pt}

\pgfmathsetlengthmacro{\SAUnit}{(\linewidth-2pt)/100}
\newcommand{\SAFont}[1]{\pgfmathsetmacro{\SALeading}{1.18*#1}\normalfont\fontfamily{ptm}\fontsize{#1}{\SALeading}\selectfont}

\definecolor{SAUI}{HTML}{E8F3FC}
\definecolor{SAUIEdge}{HTML}{5E94B9}
\definecolor{SAIO}{HTML}{EDF5E7}
\definecolor{SAIOEdge}{HTML}{779D69}
\definecolor{SAPrep}{HTML}{FFF4DE}
\definecolor{SAPrepEdge}{HTML}{B49450}
\definecolor{SACore}{HTML}{E3F3F0}
\definecolor{SACoreEdge}{HTML}{58998F}
\definecolor{SASearch}{HTML}{EFEAF9}
\definecolor{SASearchEdge}{HTML}{9580BE}
\definecolor{SAResult}{HTML}{FBECE6}
\definecolor{SAResultEdge}{HTML}{BC8875}
\definecolor{SAInk}{HTML}{22272D}
\definecolor{SALine}{HTML}{899097}
\definecolor{SAFileInk}{HTML}{646B73}
\definecolor{SACard}{HTML}{FCFCFD}
\definecolor{SAIcon}{HTML}{D9E8E1}

\tikzset{
  sa/group/.style={
    rounded corners=4pt,
    line width=0.5pt,
    anchor=north west,
    inner sep=0pt,
    outer sep=0pt
  },
  sa/card/.style={
    rounded corners=1.5pt,
    draw=SALine,
    fill=SACard,
    line width=0.35pt,
    anchor=north west,
    inner sep=0pt,
    outer sep=0pt
  },
  sa/title/.style={
    font=\SAFont{\SAClassSize}\bfseries,
    text=SAInk,
    inner sep=0pt,
    outer sep=0pt,
    align=center
  },
  sa/file/.style={
    font=\SAFont{\SAFileSize},
    text=SAFileInk,
    anchor=north west,
    inner sep=0pt,
    outer sep=0pt,
    align=left
  },
  sa/body/.style={
    font=\SAFont{\SABodySize},
    text=SAInk,
    anchor=west,
    inner sep=0pt,
    outer sep=0pt
  },
  sa/label/.style={
    font=\SAFont{\SALabelSize},
    text=SAInk,
    inner sep=0.8pt,
    align=center
  },
  sa/mark/.style={
    circle,
    draw=SAFileInk,
    fill=SAIcon,
    line width=0.3pt,
    minimum size=\SAMarkerDiameter,
    inner sep=0pt,
    outer sep=0pt,
    font=\SAFont{\SAMarkerSize}\bfseries
  },
  sa/call/.style={
    draw=SAInk,
    line width=0.45pt,
    -{Stealth[length=2.8pt,width=2.3pt]},
    rounded corners=2pt
  },
  sa/return/.style={
    sa/call,
    dash pattern=on 2pt off 1.6pt
  },
  sa/optional/.style={
    sa/call,
    dash pattern=on 0.6pt off 1.7pt
  },
  sa/contains/.style={
    draw=SAInk,
    line width=0.45pt,
    {Diamond[length=3.3pt,width=3pt]}-
  }
}

\newcommand{\SAGroup}[7]{\pgfmathsetlengthmacro{\SAWidth}{#3*\SAUnit}\node[
    sa/group,
    minimum width=\SAWidth,
    minimum height=#4pt,
    fill=#5,
    draw=#6
  ] (#1) at #2 {};
  \node[
    anchor=west,
    font=\SAFont{\SAGroupSize}\bfseries,
    text=SAInk,
    inner sep=0pt
  ] at ([xshift=5pt,yshift=-8pt]#1.north west) {#7};
}

\newcommand{\SABox}[8]{\pgfmathsetlengthmacro{\SAWidth}{#3*\SAUnit}\node[
    sa/card,
    minimum width=\SAWidth,
    minimum height=#4pt
  ] (#1) at #2 {};

  \node[sa/mark] at ([xshift=5.6pt,yshift=-6.5pt]#1.north west) {#5};
  \node[sa/title] at ([yshift=-6.5pt]#1.north) {#6};

  \draw[SALine,line width=0.3pt] ([yshift=-13pt]#1.north west) -- ([yshift=-13pt]#1.north east);

  \node[
    sa/file,
    text width=\dimexpr\SAWidth-7pt\relax
  ] (#1-file) at ([xshift=3.5pt,yshift=-16pt]#1.north west) {#7};

  \foreach \SAEntry [count=\SAIndex] in {#8}{\pgfmathsetmacro{\SARowShift}{6.2+(\SAIndex-1)*1.25*\SABodySize}\coordinate (sa-row) at ([yshift=-\SARowShift pt]#1-file.south west);
    \fill[SAFileInk] ([xshift=0.7pt]sa-row) circle[radius=0.65pt];
    \node[sa/body] at ([xshift=4pt]sa-row) {\SAEntry};
  }}

\begin{tikzpicture}[x=\SAUnit,y=-1pt]
\path[use as bounding box] (0,0) rectangle (100,344);

\SAGroup{ui}{(0,0)}{20}{119}
  {SAUI}{SAUIEdge}{User Interfaces}

\SAGroup{io}{(0,162)}{20}{154}
  {SAIO}{SAIOEdge}{I/O and Validation}

\SAGroup{prep}{(41,0)}{59}{76}
  {SAPrep}{SAPrepEdge}{Preprocessing and Initialization}

\SAGroup{core}{(28,109)}{19.6}{78}
  {SACore}{SACoreEdge}{Solver Core}

\SAGroup{search}{(61.8,80)}{38.2}{236}
  {SASearch}{SASearchEdge}{Search Engine and Algorithms}

\SAGroup{results}{(28,235)}{30}{81}
  {SAResult}{SAResultEdge}{Solution and Results}

\SABox{cli}{(1.2,20)}{17.6}{42}{M}{Command Line}
  {main.cpp; cli.hpp / .cpp}
  {parse\_command\_line(\ldots),makeConfig(\ldots)}

\SABox{batch}{(1.2,70)}{17.6}{42}{M}{Batch Runner}
  {batch.hpp / .cpp}
  {collect\_batch\_cases(\ldots),run\_batch(\ldots)}

\SABox{reader}{(1.2,182)}{17.6}{35}{M}{Instance Reader}
  {instance\_io.hpp / .cpp}
  {read\_instance(\ldots)}

\SABox{writer}{(1.2,225)}{17.6}{42}{M}{Result Writer}
  {result\_io.hpp / .cpp}
  {write\_assignment(\ldots),append\_result\_csv(\ldots)}

\SABox{checker}{(1.2,275)}{17.6}{35}{M}{Assignment Check}
  {algorithms.hpp / .cpp}
  {check\_assignment(\ldots)}

\SABox{initial}{(42.3,20)}{21}{50}{M}{Initial Bounds}
  {initial\_bounds.hpp / .cpp}
  {computeBounds(\ldots),dffBound(\ldots),restoreAssignment(\ldots)}

\SABox{initialres}{(66.5,20)}{13.8}{50}{S}{Initial Result}
  {initial\_bounds.hpp}
  {lower\_bound,incumbent,prepared}

\SABox{prepared}{(83.5,20)}{15.2}{50}{S}{Prepared Data}
  {initial\_bounds.hpp}
  {search\_instance,search\_incumbent,search\_to\_original}

\SABox{solverapi}{(29.3,129)}{17}{50}{M}{Solver API}
  {solver.hpp / .cpp}
  {solve(\ldots),verify\_gurobi\_runtime()}

\SABox{bbr}{(63.3,100)}{16.2}{42}{M}{BBR Driver}
  {bbr.hpp / .cpp}
  {runBBR(\ldots)}

\SABox{engine}{(82.5,100)}{16.2}{42}{C}{BBR Engine}
  {bbr.cpp}
  {solve(),search()}

\SABox{bound}{(63.3,152)}{16.2}{42}{C}{Bin Bound}
  {bin\_packing\_bound.hpp / .cpp}
  {solve(\ldots),lookup\_exact(\ldots)}

\SABox{conflict}{(82.5,152)}{16.2}{42}{C}{Conflict Bound}
  {conflict\_bin\_packing.hpp / .cpp}
  {solve(\ldots),quick\_lower\_bound(\ldots)}

\draw[SASearchEdge!55,line width=0.3pt] (63.3,202) -- (98.7,202);

\node[
  anchor=west,
  font=\SAFont{\SAFileSize}\itshape,
  text=SAFileInk,
  inner sep=0pt
] at (63.3,209) {Optional root strengthening (Gurobi)};

\SABox{rootcg}{(63.3,218)}{16.2}{47}{M}{Root CG}
  {root\_column\_generation.*}
  {runRootCG(\ldots)}

\SABox{master}{(82.5,218)}{16.2}{47}{C}{Root Master}
  {root\_column\_generation.cpp}
  {solve(\ldots),duals(),add\_pattern(\ldots)}

\SABox{longpricing}{(63.3,276)}{16.2}{34}{C}{{Column Pricer}}
  {root\_column\_generation.cpp}
  {solve()}

\begin{scope}[sa/title/.append style={xshift=4pt}]
\SABox{integerpricing}{(82.5,276)}{16.2}{34}{C}{{Certification Pricer}}
  {root\_column\_generation.cpp}
  {solve()}
\end{scope}

\SABox{solution}{(29.2,255)}{12.6}{54}{S}{Solution}
  {types.hpp}
  {status,assignment,lower\_bound,upper\_bound}

\SABox{statistics}{(44.2,255)}{12.6}{54}{S}{Statistics}
  {types.hpp}
  {explored\_nodes,generated\_columns,total\_seconds}

\draw[sa/call] (ui.east) -- (24,0 |- ui.east) |- (34,98) -- (34,109);
\node[sa/label,anchor=south] at (31,97) {configure and solve};

\draw[sa/call] (ui.south) -- (io.north);
\node[sa/label,anchor=west] at (10.8,156) {read / write};

\draw[sa/call] (44,109) -- (44,76);
\node[sa/label,anchor=west] at (45,92) {initial bounds};

\draw[sa/call] (core.west) -- (25.5,0 |- core.west) |- (checker.east);
\node[sa/label,anchor=south] at (23,290) {check};

\draw[sa/return] (initial.east) -- (initialres.west);
\draw[sa/contains] (initialres.east) -- (prepared.west);

\draw[sa/call] (core.east) -- (56,0 |- core.east) |- (bbr.west);
\node[sa/label,anchor=south] at (54.8,118) {calls and coordinates};

\coordinate (optionalstart) at ([yshift=8pt]core.south east);
\draw[sa/optional] (optionalstart) -- (59.5,0 |- optionalstart) |- (rootcg.west);
\node[sa/label,anchor=east] at (58.8,230) {optional};

\draw[sa/return] (38,187) -- (38,235);
\node[sa/label,anchor=west] at (39,211) {returns solution};

\draw[sa/contains] (solution.east) -- (statistics.west);

\draw[sa/call] (bbr.east) -- (engine.west);

\draw[sa/call] (engine.south) -- (90.6,147) -- (71.4,147) -- (bound.north);

\draw[sa/call] (bound.east) -- (conflict.west);

\draw[sa/call] (rootcg.east) -- (master.west);

\draw[sa/call] (rootcg.south) -- (longpricing.north);

\draw[sa/call] ([xshift=7pt]rootcg.south) -- ++(0,6) -| (integerpricing.north);

\pgfmathsetlengthmacro{\SALegendWidth}{100*\SAUnit}

\node[
  sa/group,
  draw=SALine,
  fill=white,
  minimum width=\SALegendWidth,
  minimum height=20pt
] (legend) at (0,324) {};

\def\SALegendY{334}

\tikzset{
  sa/legend text/.style={
    sa/label,
    anchor=west,
    inner sep=0pt,
    text height=4.4pt,
    text depth=1.2pt
  }
}

\node[
  sa/legend text,
  font=\SAFont{\SALabelSize}\bfseries
] at (2,\SALegendY) {Legend:};

\node[sa/mark] at (13,\SALegendY) {M};
\node[sa/legend text] at (14.2,\SALegendY) {Module};

\node[sa/mark] at (25,\SALegendY) {C};
\node[sa/legend text] at (26.2,\SALegendY) {Class};

\node[sa/mark] at (36,\SALegendY) {S};
\node[sa/legend text] at (37.2,\SALegendY) {Struct};

\draw[sa/call] (46,\SALegendY) -- (48.8,\SALegendY);
\node[sa/legend text] at (49.8,\SALegendY) {Calls};

\draw[sa/return] (58,\SALegendY) -- (60.8,\SALegendY);
\node[sa/legend text] at (61.8,\SALegendY) {Returns};

\draw[sa/contains] (72,\SALegendY) -- (74.8,\SALegendY);
\node[sa/legend text] at (75.8,\SALegendY) {Contains};

\draw[sa/optional] (86,\SALegendY) -- (88.8,\SALegendY);
\node[sa/legend text] at (89.8,\SALegendY) {Optional};

\end{tikzpicture}
\par
\endgroup

\caption{Component Interfaces and Principal Dependencies in \solver. CG: Column Generation.}
\label{fig:solver-interface}
\end{figure}

The executable used by the CLI and sequential batch launcher links to the static library \texttt{precpack\_core} and invokes the same \texttt{solve} API used by library clients. As illustrated in Figure~\ref{fig:solver-interface}, input/output (I/O) and assignment checking are separated from the search algorithms, allowing applications to reuse these routines. The Solver API coordinates initialization, optional root strengthening, and the BBR Driver. Root strengthening returns a lower-bound result without maintaining BBR states or determining the final termination status. All three problem classes share a single BBR implementation; fixed settings select applicable rules and simpler representations according to precedence weights rather than benchmark-set names. Section~\ref{sec:algorithm} specifies the corresponding applicability conditions and effort limits. Section~\ref{app:interface-identifiers} of the Online Supplement provides the C++ identifiers corresponding to the shortened names in the figure.

The public function \texttt{solve(const Instance\&, const Config\&)} accepts an \texttt{Instance} containing the problem data and a \texttt{Config} containing the algorithmic settings and resource limits. It returns a \texttt{Solution} containing the assignment, bounds, status, and statistics. The function \texttt{make\_solver\_config} constructs a \texttt{Config} with fixed algorithmic settings and the requested resource limits. These limits must be finite and positive, including when a \texttt{Config} is constructed directly.

Figure~\ref{fig:minimal-usage} illustrates a call with a 300-second time limit and a limit of 24 gibibytes (GiB; $2^{30}$ bytes) on accounted BBR memory. This memory limit is passed to the API as 24576 mebibytes (MiB; $2^{20}$ bytes). The statuses \texttt{OPTIMAL}, \texttt{TIME\_LIMIT}, and \texttt{MEMORY\_LIMIT} distinguish optimality from resource-limited termination. Invalid input and execution errors produce library exceptions or nonzero CLI exit codes.

\begin{figure}[t]
\centering
\newcommand{\FigureCodeFontSize}{8pt} \begin{lstlisting}[style=precpackcpp,basicstyle=\fontencoding{T1}\ttfamily\fontsize{\FigureCodeFontSize}{\dimexpr\FigureCodeFontSize*6/5\relax}\selectfont]
const auto problem = precpack::ProblemKind::kBppGp;
const auto instance = precpack::read_instance(
    "case.txt", std::filesystem::path{"case.graph"}, precpack::to_string(problem));
const auto config = precpack::make_solver_config(problem, 300.0, 24576);
const auto result = precpack::solve(instance, config);
std::cout << precpack::to_string(result.status)
          << " LB=" << result.lower_bound << " UB=" << result.upper_bound << '\n';
\end{lstlisting}
\caption{Core Library Call Through the Unified Interface}
\label{fig:minimal-usage}
\end{figure}

\subsection{Efficient Implementation}\label{sec:efficient-implementation}

The implementation limits memory use and avoids repeated computation. States are stored as contiguous 64-bit words, while frequently accessed fields are stored in separate arrays allocated in blocks. Unnecessary fields are omitted when the separation tuple is empty. A hash table supports state lookup, and full-state comparison prevents hash collisions from causing distinct states to be treated as equal. State lookup, grouping, reinsertion, and memory accounting are detailed in Section~\ref{app:state-storage} of the Online Supplement.
Bit masks and reusable arrays support eligibility and dominance tests without reconstructing successor relations at each branch. DFF contributions are precomputed, and each child's residual sums are obtained by subtracting the contributions of its newly assigned items from the parent's sums. Auxiliary bin-packing searches also reuse allocated storage. Further details are provided in Sections~\ref{app:dff-selection}, \ref{app:conflict-binlb}, and \ref{app:load-enumeration} of the Online Supplement.

\subsection{Reuse and Extension}\label{sec:reuse-extension}

Initialized with an instance and effort limits, \texttt{BinPackingBound} can be reused to evaluate different item subsets without invoking the complete BBR search. For a state at depth $d$, it receives the unassigned-item set, the global deadline, the available time budget, and the target $\mathrm{UB}-d$. Subject to its effort limits, it is called at the final bound stage in Section~\ref{sec:search-bounds}, after earlier bounds and memory and dominance tests have failed to prune the state. It returns a valid lower bound on the number of additional bins and completion information. BBR adds $d$ to this bound and takes the maximum with the existing state bound, pruning when the result reaches $\mathrm{UB}$. If the call is interrupted, only established bounds are retained, including any completed ordinary-bin-packing optimum; an unfinished search does not establish infeasibility. New residual bounds can be added to this sequence according to computational cost, with the same requirements for bound validity and interruption handling. New primal heuristics similarly return assignments for independent checking before the upper bound is updated. Extending the problem definition also requires establishing state sufficiency, branching completeness, and the validity of preprocessing and assignment checks.

\subsection{Execution and Reproducibility}\label{sec:software-reproducibility}\label{sec:resource-control}

Common \texttt{.txt} files store weights, capacity, and precedence arcs; SALBP-I and BPP-P assign arc weights of zero and one, respectively, whereas BPP-GP uses a paired weighted \texttt{.graph} file. Results are written to a compact comma-separated values (CSV) file and a separate assignment file that preserves intermediate empty bins. Sequential batch runs can be resumed.
CMake builds the C++20 library and executable on macOS, Linux, and Windows. Gurobi is optional and used only for root-bound strengthening and reference tests; preprocessing, primal heuristics, complete BBR, and validation remain available without it. BBR and Gurobi each use one central processing unit (CPU) thread. Fixed orders, tie-breaking rules, and seed 1 determine algorithmic choices. Wall-clock interruption can nevertheless change the work completed across machines or builds.

One global deadline applies to all solution phases. The memory limit applies to the memory tracked for BBR, not to the resident memory of the entire process; the checks before additional memory is allocated are specified in Section~\ref{app:state-storage} of the Online Supplement. Bounds and assignments are retained upon termination as described in Section~\ref{sec:reconstruction-validation}.
Our \href{https://github.com/Sunkanghong-Wang/PrecPack}{MIT-licensed repository} contains the source code and benchmark data. A separate replication archive will preserve the fixed source revision, inputs, settings, raw results, and table-generation scripts used in this paper, allowing subsequent software development without altering the published evidence. Section~\ref{app:experimental-settings} of the Online Supplement provides the computational settings.

\subsection{Verification and Regression Tests}\label{sec:software-verification}

The reusable \texttt{check\_assignment} function checks candidate and returned assignments against an instance without relying on the changing search state. Exhaustive reference solvers for small instances complement this check by testing preprocessing, dominance, bounds, and complete BBR. Conflict-aware BINLB is compared with an independent solver for bin packing with conflicts; the BPP-P flow bound is tested on targeted examples and 80 randomized instances with brute-force optima. Optional Gurobi tests compare root bounds and reference integer-programming solutions.

Regression cases cover forced empty bins, reinsertion of states reached again at a smaller depth, hash collisions, and interrupted auxiliary searches. Interface tests cover argument parsing, configurations, data handling, CSV compatibility, batch resumption, and time-limit reporting. The GitHub Actions workflow runs Gurobi-free builds, tests, and checks of the launch scripts on all three platforms.

\section{Computational Experiments}\label{sec:experiments}

We evaluate \solver\ through same-machine comparisons with source-available exact methods and comparisons with published results across SALBP-I, BPP-P, and BPP-GP. Experiments use a MacBook Pro with an Apple M4 Pro processor and a nominal memory capacity of 48 gigabytes (GB; $10^9$ bytes). The C++ implementation is compiled in Release mode with Apple Clang 17.0.0 and includes support for Gurobi 13.0.2 for root-bound strengthening. All \solver\ runs use one CPU thread, seed 1, a 24 GiB limit on the memory accounted for by BBR, and the fixed algorithm settings of Section~\ref{sec:algorithm}.
In the following tables, ``\#I'' denotes the number of instances and ``\#Opt'' the number proved optimal. Lower-bound matches reported for BM-ILS are identified separately. For \solver, ``Time/s'' denotes the average elapsed computing time over all instances, including those not proved optimal; it includes preprocessing, initialization, root strengthening when invoked, and BBR, but excludes file I/O and the time spent launching the solver. ``Gap/\%'' denotes the average of $100(\mathrm{UB}-\mathrm{LB})/\mathrm{UB}$. To retain the metric used in the SALBP-I reports, Table~\ref{tab:published-salbp-summary} instead reports relative percentage deviation (RPD), with $\mathrm{LB}$ as the denominator. Published values retain their original timing definitions and reference bounds, and the two percentages are not interchangeable. A double dash denotes unavailable data; boldface marks the best displayed value within each matched group, not a normalized performance ranking. The time limits for comparisons with published results were fixed before the experiments. Section~\ref{app:time-limits} of the Online Supplement specifies the reference methods, CPU ratings, and exceptions used to assess these limits; the ratings are not used to normalize computing times.

\subsection{Benchmark Instances}

The Otto sets generated by \citet{otto2013systematic} differ in graph structure, order strength (the proportion of item pairs ordered by precedence), and item-weight distribution. The original sets include $n\in\{20,50,100,1000\}$; additional sets with $n\in\{250,500,750\}$ were created by \citet{kramer2017batching} using the same parameters. For each size, the set contains 21 classes of 25 instances, resulting in 525 instances; we refer to these groups as Otto-20, Otto-50, and so on. The same item weights, capacities, and precedence arcs define SALBP-I with $t_{ij}=0$, BPP-P with $t_{ij}=1$, and two BPP-GP sets with precedence weights drawn uniformly from $\{0,1\}$ or $\{0,1,2,3\}$. We use these related sets to evaluate performance across problem variants and instance sizes. For SALBP-I, the original benchmark also provides a permuted Otto-50 set: each base instance is supplemented with nine random permutations of its item weights over the unchanged precedence graph, resulting in 5250 instances, including the 525 originals.

The Scholl set generated by \citet{scholl1993data} contains 269 instances obtained by varying the bin capacities of 25 base instances with 7 to 297 items. Its SALBP-I and BPP-P versions have identical item weights, capacities, and arcs, and differ only in whether every arc has weight 0 or 1 \citep{dell2012bin}. Scholl complements the systematic Otto sets with established instances of varying sizes and structures.

\subsection{Same-Machine Comparison with Source-Available Exact Solvers}

In the same-machine comparison, we evaluate the proposed solver against the three publicly available SALBP-I implementations identified in Table~\ref{tab:review-summary}: standard BBR12 \citep{sewell2012branch}, its memory-saving backtracking implementation BBR14 \citep{morrison2014application}, and \revisedbbr\ \citep{alvarezmiranda2023analysis}. All four implementations are compiled in Release mode and run sequentially on the same 525 Otto-100 instances, each using one thread and a nominal 350-second limit.
Table~\ref{tab:salbp-open-source} reports the sums of bounds across all 525 instances. ``Limit/s'' indicates the time limit per instance in seconds. ``Nodes'' reflects the average number of states generated by \solver\ and the average native node count for \revisedbbr.
Node counts from different implementations are not directly comparable and are reported only as descriptive statistics.
Final node counts are unavailable for BBR12 and BBR14. On this set, \solver\ proves optimality for 523 out of 525 instances, exceeding each of the three baselines by 66 or 67 instances, with only two instances not proved optimal within the time limit. It also attains a greater lower bound and a smaller upper bound. The average final gap is 0.008\%, compared to 0.479\% to 0.774\% for the baselines, and its average computing time is 1.79 seconds, versus recorded baseline times of 50.39 to 55.18 seconds.

\begin{table}[htbp]
\centering
\caption{Same-Machine Comparison on SALBP-I Instances with 100 Items}
\fontsize{8pt}{8pt}\selectfont
\renewcommand\arraystretch{1.0}
\tabcolsep=8.5pt
\resizebox{\textwidth}{!}{\makebox[\textwidth][c]{
\begin{tabular}{lcrlrrrrrrr}
\toprule
Instance Set & $n$ & \#I & Solver & Limit/s & \#Opt & $\sum \mathrm{LB}$ & $\sum \mathrm{UB}$ & Gap/\% & Nodes & Time/s \\
\midrule
\multirow[c]{4}{*}{Otto ($t_{ij}=0$)} & \multirow[c]{4}{*}{100} & \multirow[c]{4}{*}{525} & BBR12 & 350 & 456 & 15982 & 16117 & 0.479 & -- & 51.35 \\
 &  &  & BBR14 & 350 & 457 & 15892 & 16116 & 0.774 & -- & 55.18 \\
 &  &  & \revisedbbr & 350 & 456 & 15982 & 16117 & 0.479 & 1992763 & 50.39 \\
 &  &  & \lightgraycell \solver\ (Ours) & \lightgraycell 350 & \lightgraycell \textbf{523} & \lightgraycell \textbf{16105} & \lightgraycell \textbf{16107} & \lightgraycell \textbf{0.008} & \lightgraycell 328880 & \lightgraycell \textbf{1.79} \\
\bottomrule
\end{tabular}}}
\label{tab:salbp-open-source}
\end{table}

\subsection{Comparison with the Published Results}\label{sec:published-summary}

Tables~\ref{tab:published-salbp-summary} to~\ref{tab:published-bppgp-summary} extend the evaluation to matched benchmark groups with published results. Where available, original reports are used, retaining the longest reported limit for each method under comparison. Section~\ref{app:baseline-platforms} of the Online Supplement details the sources of the published results and the computational environments used. The time limits appear after method names in the tables. Tables~\ref{tab:published-salbp-summary} and \ref{tab:published-bppp-summary} present item count and instance count as $n$ (\#I), while each entry in Table~\ref{tab:published-bppgp-summary} provides \#Opt/Gap/Time for 525 instances. We interpret computing times alongside proof counts and reported gaps; a shorter average time under a reduced limit does not necessarily indicate greater speed, since unresolved instances are stopped earlier. Gaps computed with different reference lower bounds are unsuitable for direct comparison of feasible solution quality.

\subsubsection{SALBP-I.}

Table~\ref{tab:published-salbp-summary} compares \solver\ with SALOME97 \citep{scholl1997salome}, BBR12 \citep{sewell2012branch}, BBR14 \citep{morrison2014application}, CPLEX \citep{kramer2017batching}, the four improved methods studied by \citet{li2020comparative}, and BCP22 \citep{letelier2022timelags}. It covers Scholl, all seven Otto base sizes, and the separate permuted Otto-50 set.
Our \solver\ proves optimality for all 269 Scholl instances and all instances in the 20- and 50-item Otto groups, including the permuted Otto-50 set. At $n=100$, its 523 proofs exceed the 512 to 517 reported for the improved methods and 490 for BCP22. At $n\in\{250,500,750\}$, it proves optimality for 330 to 444 instances within 75 seconds, compared with 0 to 71 for the CPLEX runs at 300 seconds. At $n=1000$, however, our solver proves optimality for 306 instances, compared with 349 or 350 for BBR14 and the improved methods under their respective longer time limits.

\begin{table}[htbp]
\centering
\caption{Comparison with Published Results on SALBP-I}
\fontsize{8pt}{8pt}\selectfont
\renewcommand{\arraystretch}{1.0}
\setlength{\tabcolsep}{6.5pt}
\resizebox{\textwidth}{!}{\begin{tabular}{@{}l@{\hspace{12pt}}l@{}}
\toprule
\begin{tabular}[t]{rlrrr@{}}
\multicolumn{5}{c}{Scholl and Otto Base Sets, $n\le 100$} \\
\cmidrule(lr){1-5}
$n$ (\#I) & Solver (Limit/s) & \#Opt & RPD/\% & Time/s \\
\midrule
\multirow[c]{3}{*}{\shortstack{7 to 297 (269)}} & BBR12 (3600) & \textbf{269} & \textbf{0.00} & 0.43 \\
 & I-BBR20 (900) & \textbf{269} & \textbf{0.00} & -- \\
 & \lightgraycell \solver\ (350) & \lightgraycell \textbf{269} & \lightgraycell \textbf{0.00} & \lightgraycell \textbf{0.30} \\
\cmidrule(lr){1-5}
\multirow[c]{5}{*}{20 (525)} & CPLEX (300) & \textbf{525} & \textbf{0.00} & 0.09 \\
 & SALOME97 (20) & 521 & -- & -- \\
 & BBR14 (3600) & \textbf{525} & \textbf{0.00} & 0.0018 \\
 & BCP22 (3600) & \textbf{525} & -- & 0.84 \\
 & \lightgraycell \solver\ (1000) & \lightgraycell \textbf{525} & \lightgraycell \textbf{0.00} & \lightgraycell \textbf{0.0002} \\
\cmidrule(lr){1-5}
\multirow[c]{3}{*}{50 (525)} & CPLEX (300) & 482 & 0.89 & 53.85 \\
 & BCP22 (3600) & 522 & -- & -- \\
 & \lightgraycell \solver\ (1000) & \lightgraycell \textbf{525} & \lightgraycell \textbf{0.00} & \lightgraycell \textbf{0.0023} \\
\cmidrule(lr){1-5}
\multirow[c]{9}{*}{100 (525)} & CPLEX (300) & 306 & 3.85 & 133.36 \\
 & SALOME97 (70) & 355 & 2.24 & -- \\
 & BBR14 (3600) & 513 & 0.06 & -- \\
 & I-SALOME20 (900) & 512 & 0.05 & -- \\
 & I-BDP20 (900) & 514 & 0.04 & -- \\
 & I-BBR20$'$ (900) & 516 & 0.03 & -- \\
 & I-BBR20 (900) & 517 & 0.03 & -- \\
 & BCP22 (3600) & 490 & -- & -- \\
 & \lightgraycell \solver\ (350) & \lightgraycell \textbf{523} & \lightgraycell \textbf{0.01} & \lightgraycell \textbf{1.79} \\
\end{tabular}
&
\begin{tabular}[t]{@{}rlrrr}
\multicolumn{5}{c}{Otto ($n\ge 250$) and Permuted Otto-50} \\
\cmidrule(lr){1-5}
$n$ (\#I) & Solver (Limit/s) & \#Opt & RPD/\% & Time/s \\
\midrule
\multirow[c]{2}{*}{250 (525)} & CPLEX (300) & 71 & 7.44 & 266.00 \\
 & \lightgraycell \solver\ (75) & \lightgraycell \textbf{444} & \lightgraycell \textbf{0.23} & \lightgraycell \textbf{13.34} \\
\cmidrule(lr){1-5}
\multirow[c]{2}{*}{500 (525)} & CPLEX (300) & 1 & 6.94 & 300.26 \\
 & \lightgraycell \solver\ (75) & \lightgraycell \textbf{353} & \lightgraycell \textbf{0.70} & \lightgraycell \textbf{25.37} \\
\cmidrule(lr){1-5}
\multirow[c]{2}{*}{750 (525)} & CPLEX (300) & 0 & 6.55 & 293.37 \\
 & \lightgraycell \solver\ (75) & \lightgraycell \textbf{330} & \lightgraycell \textbf{0.98} & \lightgraycell \textbf{29.47} \\
\cmidrule(lr){1-5}
\multirow[c]{8}{*}{1000 (525)} & CPLEX (300) & 0 & 5.45 & 301.27 \\
 & SALOME97 (100) & 186 & -- & -- \\
 & BBR14 (3600) & \textbf{350} & 1.21 & 1100 \\
 & I-SALOME20 (900) & 349 & 0.93 & -- \\
 & I-BDP20 (900) & \textbf{350} & 0.90 & -- \\
 & I-BBR20 (900) & \textbf{350} & 0.84 & -- \\
 & I-BBR20$'$ (900) & \textbf{350} & \textbf{0.78} & -- \\
 & \lightgraycell \solver\ (350) & \lightgraycell 306 & \lightgraycell 1.00 & \lightgraycell \textbf{151.20} \\
\cmidrule(lr){1-5}
\multirow[c]{3}{*}{\shortstack{50, permuted\\(5250)}} & SALOME97 (50) & 4404 & -- & -- \\
 & BBR14 (3600) & \textbf{5250} & \textbf{0.000} & 0.21 \\
 & \lightgraycell \solver\ (1000) & \lightgraycell \textbf{5250} & \lightgraycell \textbf{0.000} & \lightgraycell \textbf{0.0027} \\
\cmidrule(lr){1-5}
\end{tabular} \\
\bottomrule
\end{tabular}}
\label{tab:published-salbp-summary}
\end{table}

\subsubsection{BPP-P.}

Table~\ref{tab:published-bppp-summary} uses the Scholl results for CPLEX and BB12 from \citet{dell2012bin}, the \enumsixteen\ results from \citet{pereira2016procedures}, the Otto CPLEX and \enumseventeen\ results from \citet{kramer2017batching}, and BCP22 results from \citet{letelier2022timelags}. For Otto, BCP22 covers $n\in\{20,50,100\}$, while CPLEX is the available published baseline for the larger sizes.
On Otto-100, \solver\ proves optimality for 523 instances, compared with 502 for \enumsixteen, 508 for \enumseventeen, and 454 for BCP22. Its average gap is smaller than those reported for \enumsixteen\ and \enumseventeen. Across the remaining Otto sizes, it proves optimal for all 20- and 50-item instances and for 278 to 426 instances in the larger groups. It matches the CPLEX proof count and gap at $n=20$ and improves both reported measures for all larger Otto sizes. On Scholl, however, \solver\ proves optimality for 262 of the 269 instances, compared with 266 for BB12 and all 269 for \enumsixteen.

\begin{table}[htbp]
\centering
\caption{Comparison with Published Results on BPP-P}
\fontsize{8pt}{8pt}\selectfont
\renewcommand{\arraystretch}{1.0}
\setlength{\tabcolsep}{8pt}
\resizebox{\textwidth}{!}{\begin{tabular}{@{}l@{\hspace{12pt}}l@{}}
\toprule
\begin{tabular}[t]{rlrrr@{}}
\multicolumn{5}{c}{Scholl and Otto Base Sets, $n\le 100$} \\
\cmidrule(lr){1-5}
$n$ (\#I) & Solver (Limit/s) & \#Opt & Gap/\% & Time/s \\
\midrule
\multirow[c]{4}{*}{\shortstack{7 to 297 (269)}} & CPLEX (7200) & 188 & 4.98 & 2289.93 \\
 & BB12 (7200) & 266 & 0.03 & 157.20 \\
 & \enumsixteen\ (7200) & \textbf{269} & \textbf{0.00} & \textbf{10.94} \\
 & \lightgraycell \solver\ (1000) & \lightgraycell 262 & \lightgraycell 0.09 & \lightgraycell 31.16 \\
\cmidrule(lr){1-5}
\multirow[c]{3}{*}{20 (525)} & CPLEX (300) & \textbf{525} & \textbf{0.00} & 0.05 \\
 & BCP22 (3600) & \textbf{525} & -- & 0.18 \\
 & \lightgraycell \solver\ (75) & \lightgraycell \textbf{525} & \lightgraycell \textbf{0.00} & \lightgraycell \textbf{0.0002} \\
\cmidrule(lr){1-5}
\multirow[c]{3}{*}{50 (525)} & CPLEX (300) & 457 & 1.22 & 55.10 \\
 & BCP22 (3600) & 501 & -- & -- \\
 & \lightgraycell \solver\ (75) & \lightgraycell \textbf{525} & \lightgraycell \textbf{0.00} & \lightgraycell \textbf{0.0028} \\
\cmidrule(lr){1-5}
\multirow[c]{5}{*}{100 (525)} & CPLEX (300) & 325 & 4.11 & 125.17 \\
 & \enumsixteen\ (7200) & 502 & 0.09 & 375.90 \\
 & \enumseventeen\ (3600) & 508 & 0.06 & 308.32 \\
 & BCP22 (3600) & 454 & -- & -- \\
 & \lightgraycell \solver\ (1000) & \lightgraycell \textbf{523} & \lightgraycell \textbf{0.01} & \lightgraycell \textbf{6.72} \\
\end{tabular}
&
\begin{tabular}[t]{@{}rlrrr}
\multicolumn{5}{c}{Otto Base Sets, $n\ge 250$} \\
\cmidrule(lr){1-5}
$n$ (\#I) & Solver (Limit/s) & \#Opt & Gap/\% & Time/s \\
\midrule
\multirow[c]{2}{*}{250 (525)} & CPLEX (300) & 154 & 7.92 & 219.69 \\
 & \lightgraycell \solver\ (75) & \lightgraycell \textbf{426} & \lightgraycell \textbf{0.31} & \lightgraycell \textbf{16.41} \\
\cmidrule(lr){1-5}
\multirow[c]{2}{*}{500 (525)} & CPLEX (300) & 32 & 8.69 & 283.00 \\
 & \lightgraycell \solver\ (75) & \lightgraycell \textbf{332} & \lightgraycell \textbf{0.82} & \lightgraycell \textbf{30.60} \\
\cmidrule(lr){1-5}
\multirow[c]{2}{*}{750 (525)} & CPLEX (300) & 24 & 7.98 & 283.31 \\
 & \lightgraycell \solver\ (75) & \lightgraycell \textbf{283} & \lightgraycell \textbf{1.10} & \lightgraycell \textbf{37.76} \\
\cmidrule(lr){1-5}
\multirow[c]{2}{*}{1000 (525)} & CPLEX (300) & 6 & 6.80 & 300.31 \\
 & \lightgraycell \solver\ (75) & \lightgraycell \textbf{278} & \lightgraycell \textbf{1.28} & \lightgraycell \textbf{39.02} \\
\cmidrule(lr){1-5}
\end{tabular} \\
\bottomrule
\end{tabular}}
\label{tab:published-bppp-summary}
\end{table}

\subsubsection{BPP-GP.}

Table~\ref{tab:published-bppgp-summary} compares \solver\ with the CPLEX and BM-ILS results reported by \citet{kramer2017batching} for both sets of precedence weights and all seven sizes. BCP22 results from \citet{letelier2022timelags} are additionally available for weights in $\{0,1\}$ and $n\in\{20,50,100\}$. Each group contains 525 instances. CPLEX and BM-ILS use a 300-second limit, BCP22 uses 3600 seconds, and \solver\ uses 75 seconds. For the BM-ILS heuristic, \#Opt counts the number of feasible solutions whose objective values match the reference lower bounds.
\solver\ proves optimality for all 20- and 50-item instances in both sets, and its \#Opt exceeds all available published baseline counts in every larger group. In particular, at $n=100$ with weights in $\{0,1\}$, it proves optimality for 521 instances, compared with 491 for BCP22. At $n=1000$, it proves optimality for 319 instances with weights in $\{0,1\}$ and 247 with weights in $\{0,1,2,3\}$, compared with 0 and 41 instances for CPLEX and 215 and 72 lower-bound matches for BM-ILS. These results extend the performance evidence to both sets of generalized precedence weights.
For weights in $\{0,1\}$ at $n=1000$, BM-ILS reports a smaller average gap of 0.63\%, compared with 1.36\% for \solver; \solver\ attains the smallest displayed gap in the other 13 groups, including ties. A larger proof count can coexist with a larger group-average gap when the unresolved instances have larger gaps.
These comparisons demonstrate the practical value of a unified exact solver for the three problems and identify the largest SALBP-I instances and Scholl BPP-P as areas for further improvement.

\begin{table}[htbp]
\centering
\caption{Comparison with Published Results on BPP-GP}
\fontsize{8pt}{8pt}\selectfont
\renewcommand{\arraystretch}{1.0}
\setlength{\tabcolsep}{2.5pt}
\resizebox{\textwidth}{!}{\begin{tabular}{llrrrrrrr}
\toprule
\multicolumn{2}{l}{Entry: \#Opt/Gap/Time} & \multicolumn{7}{c}{$n$} \\
\cmidrule(lr){3-9}
Solver (Limit/s) & $t_{ij}$ & 20 & 50 & 100 & 250 & 500 & 750 & 1000 \\
\midrule
\multirow[c]{2}{*}{CPLEX (300)} & $\{0,1\}$ & \textbf{525}/\textbf{0.00}/0.11 & 464/1.06/56.42 & 310/4.08/128.44 & 109/7.75/244.88 & 19/7.52/291.33 & 0/7.32/284.29 & 0/6.11/301.30 \\
 & $\{0,1,2,3\}$ & \textbf{525}/\textbf{0.00}/0.01 & 489/0.70/30.02 & 364/3.92/93.13 & 250/6.65/160.85 & 131/8.52/227.23 & 87/7.94/230.48 & 41/7.25/282.18 \\
\midrule
\multirow[c]{2}{*}{BM-ILS (300)} & $\{0,1\}$ & \textbf{525}/\textbf{0.00}/46.30 & 510/0.11/76.89 & 456/0.36/99.09 & 292/2.23/141.79 & 268/1.77/159.91 & 233/1.65/177.97 & 215/\textbf{0.63}/193.13 \\
 & $\{0,1,2,3\}$ & \textbf{525}/\textbf{0.00}/56.01 & 510/0.10/80.49 & 439/0.41/98.43 & 265/2.04/157.55 & 179/2.50/216.19 & 88/2.71/264.45 & 72/2.04/271.42 \\
\midrule
{}BCP22 (3600) & $\{0,1\}$ & \textbf{525}/--/0.19 & 523/--/-- & 491/--/-- & -- & -- & -- & -- \\
\midrule
\multirow[c]{2}{*}{\lightgraycell \solver\ (75)} & \lightgraycell $\{0,1\}$ & \lightgraycell \textbf{525}/\textbf{0.00}/\textbf{0.00} & \lightgraycell \textbf{525}/\textbf{0.00}/\textbf{0.00} & \lightgraycell \textbf{521}/\textbf{0.01}/\textbf{0.95} & \lightgraycell \textbf{415}/\textbf{0.36}/\textbf{17.33} & \lightgraycell \textbf{345}/\textbf{0.91}/\textbf{27.06} & \lightgraycell \textbf{335}/\textbf{1.14}/\textbf{28.68} & \lightgraycell \textbf{319}/1.36/\textbf{32.68} \\
 & \lightgraycell $\{0,1,2,3\}$ & \lightgraycell \textbf{525}/\textbf{0.00}/\textbf{0.00} & \lightgraycell \textbf{525}/\textbf{0.00}/\textbf{0.02} & \lightgraycell \textbf{520}/\textbf{0.02}/\textbf{1.32} & \lightgraycell \textbf{392}/\textbf{0.49}/\textbf{20.43} & \lightgraycell \textbf{314}/\textbf{1.19}/\textbf{33.66} & \lightgraycell \textbf{255}/\textbf{1.61}/\textbf{42.04} & \lightgraycell \textbf{247}/\textbf{1.97}/\textbf{43.79} \\
\bottomrule
\end{tabular}}
\label{tab:published-bppgp-summary}
\end{table}

\section{Conclusion and Future Research Direction}\label{sec:conclusion}

This paper studied BPP-GP, a strongly $\mathcal{NP}$-hard problem arising in numerous packing and assembly line applications with joint capacity and precedence requirements. Existing exact algorithms mainly addressed classical special cases, whereas the general problem lacked an efficient open-source exact solver. To address this gap, we developed \solver\ by extending BBR to arbitrary nonnegative precedence weights. The extensions included a state representation that retained outstanding precedence restrictions, enforced empty-bin transitions, and generalized state- and item-dominance rules. Conflict-aware residual bounds strengthened the search, while the same solution procedure specialized naturally to SALBP-I and BPP-P. Common interfaces, independent assignment checking, and reproducible batch execution make its C++ implementation reusable and verifiable, while the core solver requires no commercial software.

In same-machine, single-thread comparisons with the same time limit, \solver\ proved optimality for 523 of 525 SALBP-I benchmark instances with 100 items, compared with 456 or 457 for each of three source-available BBR implementations. Average computing times over all instances were 1.79 seconds for the proposed solver and 50.39 to 55.18 seconds for the three implementations. Further comparisons with published BPP-P and BPP-GP results showed that our solver achieved optimality for more instances and reported smaller average gaps on most matched benchmark sets.

A promising direction is to extend our open-source implementation to two-dimensional cutting stock with precedence constraints, motivated by glass production \citep{parreno2021mathematical}. A central subproblem would be packing a single bin in two dimensions, for which the exact knapsack and orthogonal-packing methods of \citet{wang2025eatkg} provide a starting point. Adapting the current dominance rules, lower bounds, and state representations to accommodate geometric feasibility and cutting-pattern requirements \citep{iori2021exact}, while maintaining validity and computational efficiency, remains challenging.

\ACKNOWLEDGMENT{The authors sincerely thank the maintainers of the public benchmark sets and source-available exact-algorithm implementations used in this study \citep{scholl1993data,dell2012bin,sewell2012branch,otto2013systematic,morrison2014application,alvarezmiranda2023analysis}. They are especially grateful to \citet{kramer2017batching} for their kind assistance and helpful responses to questions raised during the early stages of this research several years ago.}

\begingroup
\SingleSpacedXI
\putbib
\endgroup
\end{bibunit}

\newpage

\begin{APPENDICES}

\renewcommand{\thesection}{\Alph{section}}
\renewcommand{\thesubsection}{\Alph{section}.\arabic{subsection}}
\renewcommand{\thesubsubsection}{\Alph{section}.\arabic{subsection}.\arabic{subsubsection}}
\numberwithin{equation}{section}
\numberwithin{table}{section}
\renewcommand{\theequation}{\Alph{section}.\arabic{equation}}
\renewcommand{\thefigure}{\Alph{section}.\arabic{figure}}
\renewcommand{\thetable}{\Alph{section}.\arabic{table}}
\setcounter{table}{0}
\renewcommand{\theHsection}{appendix.\Alph{section}}
\renewcommand{\theHsubsection}{appendix.\Alph{section}.\arabic{subsection}}
\renewcommand{\theHequation}{appendix.\Alph{section}.\arabic{equation}}

\begin{bibunit}

\begin{center}
{\fontsize{16.5}{18}\selectfont\bfseries Online Supplement}

\vspace{6pt}

{\fontsize{17}{19}\selectfont \solver: An Efficient Open-Source Exact Solver for Bin \\ Packing with Generalized Precedence Constraints}

\vspace{3pt}

{\normalsize Sunkanghong Wang, Zhengzhong Ricky You, Roberto Baldacci,\\ Baichuan Mo, Hu Qin, Lijun Wei, Zhou Xu}

\vspace{8pt}

\end{center}

\section{Additional Technical Details}\label{app:algorithmic-details}

\subsection{DFF Candidates and Selection}\label{app:dff-selection}

This subsection describes the existing DFFs used for the BBR lower bounds in Section \ref{sec:search-bounds}. We give the finite parameter ranges, selection criteria, and evaluation procedures in \solver. The candidate functions are established DFFs, not new bounds proposed here.

Let $x\in[0,1]$ denote a normalized weight, obtained by dividing an item weight by the bin capacity; thus, item $i$ has normalized weight $w_i/C$. The first two functions are those of \citet{fekete2001newappendix}. For an integer $k\ge1$, define
\begin{equation}
\phi_k^{(1)}(x)=
\begin{cases}
x, & (k+1)x\in\mathbb{Z},\\
\lfloor(k+1)x\rfloor/k, & \text{otherwise}.
\end{cases}
\label{eq:supp-dff-one}
\end{equation}
For $\varepsilon\in[0,1/2]$, define
\begin{equation}
\phi_{\varepsilon}^{(2)}(x)=
\begin{cases}
0, & x<\varepsilon,\\
x, & \varepsilon\le x\le1-\varepsilon,\\
1, & x>1-\varepsilon.
\end{cases}
\label{eq:supp-dff-two}
\end{equation}
The third function is the improvement proposed by \citet{boschetti2003twoappendix}, in the form used by \citet{dell2012binappendix}. For $\varepsilon\in(0,1/2]$, it is
\begin{equation}
\phi_{\varepsilon}^{(3)}(x)=
\begin{cases}
0, & x<\varepsilon,\\
\dfrac{\lfloor x/\varepsilon\rfloor}{\lfloor1/\varepsilon\rfloor}, & \varepsilon\le x\le1/2,\\[3pt]
1-\dfrac{\lfloor(1-x)/\varepsilon\rfloor}{\lfloor1/\varepsilon\rfloor}, & x>1/2.
\end{cases}
\label{eq:supp-dff-three}
\end{equation}
Compositions of DFFs are also dual feasible \citep{fekete2001newappendix}; the compositions used here apply $\phi_k^{(1)}$ first and then $\phi_{\varepsilon}^{(2)}$, as in \citet{pereira2016proceduresappendix}.

The base candidates comprise $\phi_k^{(1)}$ for $k\in\{1,\ldots,20\}$ and $\phi_{\varepsilon}^{(2)}$ for $\varepsilon\in\mathcal{E}_{\mathrm{base}}$, where $\mathcal{E}_{\mathrm{base}}=(\{\lfloor C/2\rfloor/C\}\cup\{w_i/C:i\in\mathcal{I},\ 2w_i\le C\})\setminus\{0\}$. This uses the small integer range and item-weight thresholds discussed by \citet{dell2012binappendix}, with the half-capacity threshold rounded down in the integer implementation. The candidates are ranked by the rule below, and at most seven are retained. Thus, the base parameter rule is fixed, but the selected functions depend on the item weights and capacity.

When every precedence weight is zero or every precedence weight is one, the candidate set expands to include $\phi_{\varepsilon}^{(2)}$ and $\phi_{\varepsilon}^{(2)}\circ\phi_k^{(1)}$ for $k\in\{1,\ldots,100\}$ and $\varepsilon\in\mathcal{E}_{\mathrm{ext}}=\{0,1/2\}\cup\{w_i/C:i\in\mathcal{I},\ 2w_i\le C\}$, following the parameter range and composition order of \citet{pereira2016proceduresappendix}. Since $\phi_0^{(2)}(x)=x$, this set also includes the uncomposed $\phi_k^{(1)}$. It further includes $\phi_{\varepsilon}^{(3)}$ for $\varepsilon\in\{1/1000,2/1000,\ldots,500/1000\}$, as used by \citet{dell2012binappendix}. These functions are valid for arbitrary precedence weights; limiting the expansion to the specified cases limits the number of DFF sums maintained for each search state without imposing a validity condition.

For ranking and evaluation, a candidate $\phi$ is represented by a positive integer capacity $c_{\phi}$ and nonnegative integer item contributions $v_{\phi i}$ satisfying $v_{\phi i}/c_{\phi}=\phi(w_i/C)$ for each $i\in\mathcal{I}$. The base representations are defined with $c_{\phi}=kC$ for $\phi_k^{(1)}$ and $c_{\phi}=C$ for $\phi_{\varepsilon}^{(2)}$. Expanded candidates use integer representations divided by the greatest common divisor of the capacity and all contributions; for $\phi_{\varepsilon}^{(3)}$, the pre-division capacity is $\lfloor1/\varepsilon\rfloor$. Within each set, duplicate representations of capacity and item contributions are removed. Define $s_{\phi}=\sum_{i\in\mathcal{I}}v_{\phi i}/c_{\phi}$. Candidates are ranked first by decreasing $\lceil s_{\phi}\rceil$, followed by decreasing $s_{\phi}$, and then by increasing $c_{\phi}$. Any remaining ties are resolved by the generation order for base candidates and the lexicographic contribution order for expanded candidates.

If $r_{\mathrm{base}}\le7$ base candidates are retained, the best $8+r_{\mathrm{base}}$ expanded candidates are considered in ranked order. Representations already present in the retained set are skipped, and at most eight additional candidates are appended. Consequently, $\mathcal{T}_{\mathrm{DFF}}$ contains at most 15 retained DFFs in integer form. Their contributions are computed once per BBR search and are neither regenerated nor reranked at each state. For a child at depth $d$, the residual sum is obtained by subtracting the contributions of its newly assigned items, and the bound is evaluated as $d+\lceil\sum_{i\in\mathcal{I}^{\mathrm{U}}}v_{\phi i}/c_{\phi}\rceil$ using integer accumulation and ceiling division. The short BBR search in Section \ref{sec:search-initialization} uses only the base selection.

Residual DFF sums use 32-bit unsigned entries when the precomputed root totals establish that all residuals are representable within this range; otherwise, 64-bit entries are retained. Because assigning items only involves subtracting nonnegative contributions, this storage choice preserves all bound values.

\subsection{Conflict-Aware BINLB}\label{app:conflict-binlb}

The conflict-aware extension of ordinary BINLB \citep{sewell2012branchappendix} solves the relaxation defined in Section \ref{sec:search-bounds}. At a BBR state $S$ with depth $d$, its primary function is to determine whether $\mathcal{I}^{\mathrm{U}}$ can be packed into at most $k=\mathrm{UB}-d-1$ bins. If infeasibility is proved, the BBR state is pruned without needing the exact relaxation optimum. A feasible auxiliary packing only verifies the feasibility of this relaxation; it is not used as a BPP-GP incumbent since bin order and precedence distances have been excluded.

The conflict graph $\mathcal{G}^{\mathrm{c}}$ remains valid for this residual relaxation. Write $i\rightsquigarrow j$ if item $j$ is reachable from item $i$ in the precedence graph, and recall that $\mathcal{I}_{ij}$ includes both endpoints and all intermediate items on directed paths from $i$ to $j$. A path with positive weight prevents its endpoints from sharing a bin. If $i\rightsquigarrow j$ and $W(\mathcal{I}_{ij})>C$, placing both endpoints in a single bin requires $b_i\le b_s\le b_j=b_i$ for every $s\in\mathcal{I}_{ij}$, violating capacity. Consequently, every conflict edge is valid. Any sequence of future loads in $\mathscr{C}(S)$, once bin order and empty positions are ignored, leads to a capacity-feasible packing of $\mathcal{I}^{\mathrm{U}}$ that respects these conflicts. Therefore, $d+z_{\mathrm{BPPC}}(\mathcal{I}^{\mathrm{U}})\le z(S,d)$, while $z_{\mathrm{BP}}(\mathcal{I}^{\mathrm{U}})\le z_{\mathrm{BPPC}}(\mathcal{I}^{\mathrm{U}})$ since the ordinary relaxation also eliminates conflicts.

Within this test, a recursive call considers a remaining item subset $\widehat{\mathcal{I}}$ and a remaining bin limit $k$. The empty subset is feasible, whereas a nonempty subset is infeasible when $k\le0$. A subset with $|\widehat{\mathcal{I}}|\le k$ is feasible when one item is assigned per bin. Otherwise, remembered bounds are checked first. Capacity, the two DFFs $\phi_1^{(1)}$ and $\phi_2^{(1)}$ from Section \ref{app:dff-selection}, and a greedy clique provide inexpensive lower bounds. The clique is constructed by scanning items in the order of nonincreasing degree in the fixed conflict graph, followed by nonincreasing weight, and then increasing identifier, including an item only if it conflicts with every item already selected. A remembered optimum of the ordinary BINLB for the same subset can further increase the lower bound. If any of these lower bounds exceeds $k$, the call is infeasible. Otherwise, a conflict-feasible best-fit packing is constructed; if it uses at most $k$ bins, the call is feasible.

If neither test resolves the call, an anchor item is selected based on nonincreasing weight, followed by nonincreasing degree in the fixed conflict graph, and then increasing identifier. The subsequent bin must include this item. Remaining items are considered in the same order using include/exclude branching, subject to capacity and pairwise compatibility constraints. Only maximal subsets containing the anchor are kept. This approach preserves an optimal auxiliary packing because the anchor's bin can be positioned first, since the bin order is irrelevant, and a nonmaximal bin can be expanded by moving compatible items from other bins without increasing the total number of bins. A candidate load $\mathcal{L}$ must satisfy
\begin{equation}
W(\mathcal{L})\ge\max\{0,W(\widehat{\mathcal{I}})-(k-1)C\},
\label{eq:supp-binlb-minimum-load}
\end{equation}
because the remaining $k-1$ bins have total capacity $(k-1)C$. During subset enumeration, a branch is discarded if the weight already selected plus the total weight of all undecided items is below this threshold. Including items that may later prove incompatible only overestimates the attainable weight, thereby preserving this pruning test's validity. For every retained load, the inexpensive bounds are checked on $\widehat{\mathcal{I}}\setminus\mathcal{L}$ before recursively testing that subset with limit $k-1$. One feasible child suffices to answer the call; infeasibility requires every possible child to be ruled out.
The auxiliary memory stores the lower and upper bounds for each subset of items, and complete subset comparisons are performed after hashing. When a feasible test at limit $k$ is performed, the upper bound is updated to at most $k$; conversely, if a test proves infeasible, the lower bound is raised to at least $k+1$. Later calls with different bin limits can therefore reuse the same records. The ordinary BINLB memory is queried only for completed ordinary optima. Conflict and subset masks are updated via bitwise operations, while total weights and DFF sums are adjusted by subtraction when a load is fixed.

The stage is considered only when at most 400 items remain unassigned and is limited to 50 generated loads per auxiliary state, 1000000 search nodes per call, 200000 states stored per relaxation, and a cumulative budget of 0.1 seconds. Within this budget, each conflict-aware test is further limited to 0.005 seconds and 50000 search nodes. These limits restrict the effort spent outside the main BBR search, not the validity of its bounds.

An interrupted recursive call returns no feasibility or infeasibility conclusion. Bounds and decisions already completed for other subsets remain valid. If a call is halted before completing its intended test, further conflict-aware calls for that BBR search are disabled when the ordinary phase has completed; otherwise, all subsequent BINLB calls are disabled. The main BBR search proceeds using the remaining bounds. This avoids repeatedly spending the auxiliary-search allowance on unfinished tests and does not discard any BBR branch based on an unproven result.

\subsection{Integer Computation of the Root Lower Bound}\label{app:root-integer-bound}

This subsection specifies the integer calculation used in Section \ref{sec:root-strengthening}, following the fixed-point approach of \citet{baldacci2024numericallyappendix}. Let $\pi_i\ge0$ and $\mu\ge0$ be the floating-point dual values with the sign convention defined there, after normalization to the objective $h$; negative values caused by LP tolerances are replaced by zero. The candidate integer scale is
\begin{equation}
\Delta=\left\lfloor\frac{2^{62}}{\max\{1,\mu+\sum_{i\in\mathcal{I}}\pi_i\}}\right\rfloor.
\label{eq:supp-root-scale}
\end{equation}
The scaled values are truncated toward zero, giving $\widehat{\pi}_i=\lfloor\Delta\pi_i\rfloor$ for every $i\in\mathcal{I}$ and $\widehat{\mu}=\lfloor\Delta\mu\rfloor$. Conversion and accumulation are checked for overflow, and the scale is reduced if necessary until $\widehat{\mu}+\sum_{i\in\mathcal{I}}\widehat{\pi}_i\le2^{62}$. All subsequent profit sums and DP values fit within the checked integer range. A positive representable scale is required before the calculation proceeds.

Column generation and final bound calculation have distinct stopping conditions. In column generation, integer pattern profits are compared against a scaled threshold to identify improving patterns. The heuristic pricing procedure limits search effort and repeated item use across generated patterns. The scale used for column generation may be smaller than $\Delta$ to match the numerical resolution of the restricted LP. If the pricing search finishes without returning a new pattern, a separate integer pricing call maximizes $\sum_{i\in\mathcal{I}}\widehat{\pi}_i\xi_i$ over all patterns $\mathscr{P}$ using the verified scale $\Delta$. This step excludes neither existing patterns nor those below $\widehat{\mu}$ and does not apply the heuristic procedure's search limit or diversification restrictions. The calculation must establish the maximum pattern profit, not merely find an improving pattern. The DP and BB comparisons use integer arithmetic, and time constraints prevent acceptance of a new bound if interrupted.

Let $V^\star(\widehat{\boldsymbol{\pi}})$ be the proven maximum integer profit and define $D=\max\{\Delta,V^\star(\widehat{\boldsymbol{\pi}})\}$. The resulting lower bound is
\begin{equation}
B_{\mathrm{root}}=\left\lceil\frac{\sum_{i\in\mathcal{I}}\widehat{\pi}_i}{D}\right\rceil.
\label{eq:supp-root-integer-bound}
\end{equation}
For every pattern $p\in\mathscr{P}$, $\sum_{i\in\mathcal{I}}a_{ip}\widehat{\pi}_i/D\le1$. Thus, any packing with $h$ bins has $\sum_{i\in\mathcal{I}}\widehat{\pi}_i/D\le h$: each nonempty bin contributes at most one, and intermediate empty bins only increase $h$. Equation \eqref{eq:supp-root-integer-bound} therefore follows from the integrality of the objective. This argument depends on the resulting nonnegative integer profits and the proven pricing maximum, not on exact dual feasibility of the floating-point LP solution. In particular, it does not require $V^\star(\widehat{\boldsymbol{\pi}})\le\widehat{\mu}$, because the denominator rescales all pattern profits to at most one.

The ratio in Equation \eqref{eq:supp-root-integer-bound} is rounded upward by integer division, without conversion back to floating point, and $\mathrm{LB}$ is replaced by $\max\{\mathrm{LB},B_{\mathrm{root}}\}$. The restricted LP objective is not rounded to obtain a pruning bound. If the integer pricing optimum has not been proved within the root-stage budget, the previously established lower bound is retained.

\subsection{Maximal-Load Enumeration}\label{app:load-enumeration}

The branching set in Section \ref{sec:maximal-load} is generated through DFS over item subsets. At a BBR state $S$, enumeration begins with $\mathcal{L}=\varnothing$ and residual capacity $C$. An unassigned item is excluded from the current bin if it is part of $\mathcal{D}_1$ or has an unassigned predecessor linked by a positive-weight arc. For all other items, the number of immediate predecessors linked by zero-weight arcs that are outside $\mathcal{I}^{\mathrm{A}}\cup\mathcal{L}$ is tracked. An item becomes eligible when this count reaches zero. Eligible items are ordered lexicographically by dominance count (the number of extended Jackson and additional subgraph-isomorphism relations favoring the item), positional weight (its weight plus the total weight of its transitive successors), item weight, and number of transitive successors, all arranged in nonincreasing order. Remaining ties are resolved by increasing stable item identifiers.
The recursive enumeration selects a capacity-feasible eligible item, inserts it into $\mathcal{L}$, and decreases the predecessor counts of its immediate successors linked by zero-weight arcs. Newly eligible successors are added to the end of the list of eligible items in the same order of priority. Recursion continues after the selected item's position, allowing newly eligible items to join the current bin while avoiding permutations of items that were already available. On return, $\mathcal{L}$, the predecessor counts, residual capacity, and accumulated DFF contributions are restored. Successors linked by positive-weight arcs are never made eligible within the current bin; their restrictions are handled when the child BBR state is formed.
Reaching the end of the current recursive extension does not establish maximality, as an earlier item may have been skipped. Before retaining $\mathcal{L}$, the algorithm checks the entire current eligible list for an unselected item that still fits. If such an item exists, the load is discarded as nonmaximal. If no such item exists, then no strictly superset is feasible. In the acyclic precedence graph, any feasible extension would contain a first additional item whose predecessors linked by zero-weight arcs are already in $\mathcal{I}^{\mathrm{A}}\cup\mathcal{L}$. That item would pass the eligibility and capacity checks and would therefore have been found by the scan. This final check implements the maximality requirement of Equation \eqref{eq:branching-load-set}.

Each maximal load is then tested against the item-replacement rules of Section \ref{sec:separation-dominance}, including the classical rules under their stated conditions, before generating a child state. If no item is initially eligible, the empty load produces the forced-empty-bin transition. An empty load is not retained merely because dominance eliminates nonempty loads. The same enumeration procedure handles arbitrary nonnegative precedence weights; zero-weight arcs permit eligibility to change within a bin, whereas positive-weight arcs defer successors to later bins.

Eligibility is recorded in flat arrays containing byte-sized status values and integer predecessor counts; bit masks represent assigned and selected subsets. For heterogeneous precedence weights, Proposition \ref{prop:labeled-item-dominance} is evaluated using precomputed masks for direct successors, positive-weight successors, and precedence-weight thresholds. If all precedence weights are set to 1, the subgraph-isomorphism search is capped at 100000 nodes per item pair and 2000000 nodes in total, with periodic deadline checks. Only proved relations are retained, and replacement capacity, along with state-specific eligibility, is checked during enumeration.

\subsection{State-Dependent Precedence Bounds}\label{app:state-dependent-bounds}

Consider a BBR state $S$ at depth $d$ with $\mathcal{I}^{\mathrm{U}}\neq\varnothing$. The bounds below extend the root head and tail constructions of \citet{dell2012binappendix}, \citet{pereira2016proceduresappendix}, and \citet{kramer2017batchingappendix} by incorporating restrictions imposed by already assigned items. The quantities $\rho_i(S)$, $e_i$, and $\eta_i$ are those defined in Section \ref{sec:search-bounds}.

In Equations \eqref{eq:residual-offsets}, $\rho_i(S)$ counts the consecutive future bins in which item $i$ remains unavailable, while predecessor terms propagate these constraints through the unassigned items. Induction in topological order shows that $e_i$ is a valid earliest offset, and reverse-topological induction shows that $\eta_i$ is a necessary tail after item $i$. Consequently, each future-load sequence in $\mathscr{C}(S)$ requires at least $e_i+\eta_i+1$ additional bins for every $i\in\mathcal{I}^{\mathrm{U}}$, establishing $\mathrm{LB}_{\mathrm{path}}\le z(S,d)$. If all precedence weights are one, the separation tuple is empty, simplifying this to the ordinary residual longest-path bound for BPP-P.

Let $u_1,\ldots,u_{|\mathcal{I}^{\mathrm{U}}|}$ and $v_1,\ldots,v_{|\mathcal{I}^{\mathrm{U}}|}$ denote the unassigned items ordered by nonincreasing $\eta_i$ and nonincreasing $e_i$, respectively. The state-dependent one-machine bounds are
\begin{equation}
\mathrm{LB}_{\mathrm{mach}}^{+}=d+\max_{1\le k\le|\mathcal{I}^{\mathrm{U}}|}\left\{\eta_{u_k}+\left\lceil\frac{\sum_{\ell=1}^{k}w_{u_\ell}}{C}\right\rceil\right\},\qquad \mathrm{LB}_{\mathrm{mach}}^{-}=d+\max_{1\le k\le|\mathcal{I}^{\mathrm{U}}|}\left\{e_{v_k}+\left\lceil\frac{\sum_{\ell=1}^{k}w_{v_\ell}}{C}\right\rceil\right\}.
\label{eq:supp-machine-bounds}
\end{equation}
For the forward bound, the first $k$ items in the $u$ order require at least $\lceil\sum_{\ell=1}^{k}w_{u_\ell}/C\rceil$ bins, with the latest scheduled of these items retaining a tail of at least $\eta_{u_k}$. For the reverse bound, none of the first $k$ items in the $v$ order can occur before offset $e_{v_k}$, beyond which their total weight requires at least $\lceil\sum_{\ell=1}^{k}w_{v_\ell}/C\rceil$ bins. Therefore, both expressions in Equation \eqref{eq:supp-machine-bounds} provide a lower bound for $z(S,d)$.

For each $i\in\mathcal{I}^{\mathrm{U}}$, define the residual predecessor and successor closures as $\mathcal{I}_i^{-}=\{i\}\cup\{j\in\mathcal{I}^{\mathrm{U}}:j\rightsquigarrow i\}$ and $\mathcal{I}_i^{+}=\{i\}\cup\{j\in\mathcal{I}^{\mathrm{U}}:i\rightsquigarrow j\}$, respectively. The state-dependent closure bound is
\begin{equation}
\mathrm{LB}_{\mathrm{cl}}=d+\max_{i\in\mathcal{I}^{\mathrm{U}}}\left\{\max\left(\left\lceil\frac{W(\mathcal{I}_i^{-})}{C}\right\rceil,e_i+1\right)+\max\left(\left\lceil\frac{W(\mathcal{I}_i^{+})}{C}\right\rceil,\eta_i+1\right)-1\right\}.
\label{eq:supp-closure-bound}
\end{equation}
All items in $\mathcal{I}_i^{-}$ must appear no later than the bin containing $i$, ensuring that the prefix of the future-load sequence through that bin has length at least both $\lceil W(\mathcal{I}_i^{-})/C\rceil$ and $e_i+1$. Correspondingly, the suffix starting with that bin has length at least both $\lceil W(\mathcal{I}_i^{+})/C\rceil$ and $\eta_i+1$. These two bin intervals overlap only at the bin containing $i$. By adding their required lengths and subtracting one, Equation \eqref{eq:supp-closure-bound} is established as a lower bound on $z(S,d)$ once the $d$ fixed bins are included.

\subsection{State Storage and Memory Accounting}\label{app:state-storage}

For separation-tuple dominance, records are grouped by assigned-item set to avoid comparisons between unrelated states. Before a one-item-superset lookup at depth $d$, item $i$ is omitted if one of its transitive predecessors is unassigned or if $W(\mathcal{I}^{\mathrm{A}})+w_i>dC$, because no state with assigned-item set $\mathcal{I}^{\mathrm{A}}\cup\{i\}$ can have been reached within the first $d$ bins. When $\tau=0$, the exact-state table is queried directly; otherwise, the separation tuples associated with the enlarged assigned set are compared.
Depths, parent identifiers, and lower bounds are stored in separate arrays, with memory allocated in contiguous blocks. Queue entries consist of a state identifier and a version number. If a state is reached at a smaller depth, its stored information and version are updated, and it is reinserted; outdated queue entries are ignored. States with an empty separation tuple omit the link used to group identical assigned sets, and when all precedence weights are zero, the duplicate assigned-set hash is also omitted.

The memory tracked by BBR includes precomputed BBR data, state storage, lookup tables, queues, temporary arrays, and auxiliary bounds. Before enlarging a state array, hash table, or queue, both the existing allocation and the storage required for any replacement are taken into account. If the required storage exceeds the limit, the search stops without discarding remembered states. Precomputed predecessor and successor masks support closure bounds, while dense masks and preallocated temporary arrays are reused by the auxiliary searches discussed in Section \ref{app:conflict-binlb}.

\subsection{C++ Identifiers in the Interface Diagram}\label{app:interface-identifiers}

To improve readability, Figure~\ref{fig:solver-interface} abbreviates selected C++ type and function names. Table~\ref{tab:interface-identifiers} lists their original identifiers; all other displayed type, function, and field names are unchanged. The arrows show selected component dependencies rather than execution order or a complete call graph.
File labels omit directory paths; \mbox{\texttt{.hpp / .cpp}} and \texttt{.*} denote the corresponding header/source pair. Command Line groups \texttt{main.cpp} and the \texttt{cli} module.

\begin{table}[!htbp]
\centering
\caption{Display Names and C++ Identifiers in Figure~\ref{fig:solver-interface}}
\fontsize{9pt}{10pt}\selectfont
\renewcommand\arraystretch{1.0}
\tabcolsep=12px
\resizebox{\textwidth}{!}{\makebox[\textwidth][c]{
\begin{tabular}{ll}
\toprule
Display Name & C++ Identifier \\
\midrule
\multicolumn{2}{l}{\textit{Type names}} \\
Prepared Data & \texttt{PreparedInstance} \\
Initial Result & \texttt{InitialBoundsResult} \\
BBR Engine & \texttt{BbrEngine} \\
Bin Bound & \texttt{BinPackingBound} \\
Conflict Bound & \texttt{ConflictBinPackingEngine} \\
Root Master & \texttt{PositionFreeMaster} \\
{Column Pricer} & \texttt{LongPricingSearch} \\
{Certification Pricer} & \texttt{IntegerPricingSearch} \\
\midrule
\multicolumn{2}{l}{\textit{Function names}} \\
\texttt{makeConfig} & \texttt{make\_command\_line\_solver\_config} \\
\texttt{computeBounds} & \texttt{compute\_initial\_bounds} \\
\texttt{dffBound} & \texttt{compute\_initial\_dff\_lower\_bound} \\
\texttt{restoreAssignment} & \texttt{map\_prepared\_assignment\_to\_original} \\
\texttt{runBBR} & \texttt{run\_branch\_bound\_remember} \\
\texttt{runRootCG} & \texttt{run\_position\_free\_root\_column\_generation} \\
\bottomrule
\end{tabular}}}
\label{tab:interface-identifiers}
\end{table}

The Column Pricer searches for new, improving patterns, whereas the Certification Pricer computes the maximum scaled profit over the full pattern set for lower-bound calculation. Both use 64-bit integer profits, but their scaling factors and stopping conditions may differ, as detailed in Section~\ref{app:root-integer-bound}.

\section{Proofs}\label{app:proofs}

\subsection{Proof of Proposition \ref{prop:state-sufficiency}}\label{app:proof-state}

Let $P$ and $\widetilde{P}$ be feasible partial packings satisfying $S_P=S_{\widetilde{P}}=S=(\mathcal{I}^{\mathrm{A}},\mathscr{D})$. By definition, their assigned sets are closed under predecessors, so no precedence arc exists from an unassigned item to an assigned item. They share the same unassigned items and have identical fixed weights, capacity, and precedence arcs. A subset $\mathcal{L}\subseteq\mathcal{I}^{\mathrm{U}}$ is feasible for the next bin if and only if $\sum_{i\in\mathcal{L}}w_i\le C$, $\mathcal{L}\cap\mathcal{D}_1=\varnothing$, each predecessor linked by a positive-weight arc to an item in $\mathcal{L}$ is in $\mathcal{I}^{\mathrm{A}}$, and each predecessor linked by a zero-weight arc is in $\mathcal{I}^{\mathrm{A}}\cup\mathcal{L}$. The condition on $\mathcal{D}_1$ enforces the remaining distances from already assigned predecessors, while other precedence conditions prevent placing an item before an unassigned predecessor. Therefore, the two partial packings admit exactly the same next loads, including the empty load, while items remain.

For either partial packing, advancing from its own depth $d$ to $d+1$ modifies the condition on an earlier assigned predecessor to $b_i+t_{ij}>d+1+r$, represented by $\mathcal{D}_{r+1}$. For a newly assigned predecessor, $b_i=d+1$ applies, so the condition becomes $t_{ij}>r$. Removing all assigned items results in precisely Equation \eqref{eq:separation-tuple-update}. Thus, a common next load generates identical child states, regardless of the two absolute depths. Induction over any finite sequence of future loads yields the same feasibility decisions and assigned sets for both partial packings. Consequently, the sequence assigns all remaining items and concludes with their final assignment from one partial packing if and only if it does so from the other. This includes the empty sequence when no items remain. As a result, $\mathscr{C}(P)=\mathscr{C}(\widetilde{P})$. \hfill\Halmos

\subsection{Proof of Proposition \ref{prop:maximal-load}}\label{app:proof-maximal-load}

Let $S$ be a state satisfying the proposition's assumptions. It admits a finite feasible sequence of future loads: the unassigned items can be placed individually in topological order, inserting finitely many empty bins as required by the precedence weights. This is possible because every item fits in a bin, and the assigned set is closed under predecessors. Hence $\mathscr{C}(S)\neq\varnothing$, and a minimum-length sequence $\boldsymbol{\mathcal{L}}=(\mathcal{L}_1,\ldots,\mathcal{L}_k)\in\mathscr{C}(S)$ exists, with $k\ge1$. If $\mathscr{F}(S)=\varnothing$, then $\mathcal{L}_1$ must be empty, because any nonempty first load would belong to $\mathscr{F}(S)$. Hence $\mathcal{L}_1=\varnothing\in\mathscr{L}(S)$.

Suppose instead that $\mathscr{F}(S)\neq\varnothing$. Since $\mathscr{F}(S)$ is finite, there exists a load $\widehat{\mathcal{L}}\in\operatorname{Max}_{\subseteq}\mathscr{F}(S)$ satisfying $\mathcal{L}_1\subseteq\widehat{\mathcal{L}}$, where the inclusion also applies when $\mathcal{L}_1=\varnothing$. Move every item in $\widehat{\mathcal{L}}\setminus\mathcal{L}_1$ from its unique later load to the first load, while retaining all intermediate bin positions. The new first load is feasible by the definition of $\widehat{\mathcal{L}}$, and every subsequent load remains capacity-feasible because it only loses items. Incoming constraints of moved items, including constraints between moved items, are satisfied by the feasibility of $\widehat{\mathcal{L}}$. Moving an item earlier cannot violate an outgoing constraint whose successor remains in a later bin. After removing any trailing empty loads, the resulting sequence belongs to $\mathscr{C}(S)$, is no longer than $\boldsymbol{\mathcal{L}}$, and begins with $\widehat{\mathcal{L}}\in\mathscr{L}(S)$. Since $\boldsymbol{\mathcal{L}}$ has minimum length, the resulting sequence also has minimum length. Thus, in either case, some minimum-length sequence in $\mathscr{C}(S)$ begins with a load in $\mathscr{L}(S)$, proving the proposition. \hfill\Halmos

\subsection{Proof of Proposition \ref{prop:tuple-dominance}}\label{app:proof-tuple}

Consider any $\boldsymbol{\mathcal{L}}'\in\mathscr{C}(S')$. Because the assigned-item sets are equal, $S$ and $S'$ have the same unassigned items and predecessor relations. If the sequence is nonempty, its first load is feasible from $S$: it avoids $\mathcal{D}'_1$ and hence $\mathcal{D}_1$, and all capacity and predecessor-closure conditions are unchanged. For any load feasible from both states, Equation \eqref{eq:separation-tuple-update} is monotone under componentwise inclusion: shifting $\mathcal{D}_{r+1}\subseteq\mathcal{D}'_{r+1}$, adding the same successor set generated by the load, and removing the same assigned items preserve inclusion in every component of the child tuple. Induction over $\boldsymbol{\mathcal{L}}'$ therefore shows that the same sequence belongs to $\mathscr{C}(S)$. The conclusion also holds for the empty sequence, since then neither state has unassigned items. If the sequence contains $k$ loads, the final bin counts satisfy $d+k\le d'+k$. Hence $S$ dominates $S'$. \hfill\Halmos

\subsection{Proof of Proposition \ref{prop:superset-dominance}}\label{app:proof-superset}

Consider any $\boldsymbol{\mathcal{L}}\in\mathscr{C}(S)$. Item $i$ occurs in exactly one load of this sequence. Delete $i$ from that load, initially retaining all bin positions. This deletion preserves capacity. Initially, $S^{+}$ has the same assigned items as $S$ together with $i$, and $\mathcal{D}^{+}_r\subseteq\mathcal{D}_r$ for every $r=1,\ldots,\tau$. The larger assigned set can only relax predecessor-membership conditions; any remaining distance restriction imposed by $i$ is already represented in $\mathscr{D}^{+}$. Therefore, the first original load with $i$ omitted satisfies the next-bin feasibility conditions at $S^{+}$. Before the original load containing $i$, the two sequences assign the same loads, their assigned sets differ only by $i$, and Equation \eqref{eq:separation-tuple-update} preserves componentwise tuple inclusion. At the load containing $i$, the transition from $S$ adds the restrictions generated by $i$, whereas the transition from $S^{+}$ omits them; the inclusion is preserved, and the assigned sets then coincide. Induction proves the capacity and precedence conditions through every remaining load, including empty loads. After removing any trailing empty loads and retaining intermediate empty bins, the resulting sequence $\widetilde{\boldsymbol{\mathcal{L}}}$ belongs to $\mathscr{C}(S^{+})$ and satisfies $|\widetilde{\boldsymbol{\mathcal{L}}}|\le|\boldsymbol{\mathcal{L}}|$. Since $d^{+}\le d$, its final objective is no larger than that of the original sequence. Thus, $S^{+}$ dominates $S$. \hfill\Halmos

\subsection{Proof of Proposition \ref{prop:labeled-item-dominance}}\label{app:proof-labeled-item}

Because the set of feasible next loads is finite and $\mathcal{L}'$ is feasible, fix a feasible maximal next load $\widehat{\mathcal{L}}\supseteq\mathcal{L}'$. This choice is made independently of the future-load sequence. Let $\mathscr{C}(S;\mathcal{L})$ denote the sequences in $\mathscr{C}(S)$ whose first load is $\mathcal{L}$, and consider any $\boldsymbol{\mathcal{L}}\in\mathscr{C}(S;\mathcal{L})$. In this sequence, item $i$ occurs in a later bin $b_i>d+1$. Replace $j$ by $i$ in the first load and replace $i$ by $j$ in its original bin $b_i$. The first replacement load is feasible by assumption; the later bin remains capacity feasible because $w_j\le w_i$. No precedence relation joins $i$ and $j$. Moving $i$ earlier relaxes its outgoing constraints, and the feasibility of $\mathcal{L}'$ verifies its incoming constraints at the new position. Moving $j$ later relaxes its incoming constraints. For any direct successor $s$ of $j$, $\theta_{is}\ge\theta_{js}\ge0$ implies $(i,s)\in\mathcal{A}$ and $t_{is}\ge t_{js}$. Thus, the original inequality $b_s-b_i\ge t_{is}$ implies that $j$'s outgoing constraint is satisfied after $j$ is moved to that bin. All other precedence constraints are unchanged, so the swapped sequence is feasible.

Next, move every item in $\widehat{\mathcal{L}}\setminus\mathcal{L}'$ from its later bin into the first bin. The first bin remains feasible by the choice of $\widehat{\mathcal{L}}$; later bins only lose weight. Arcs with both endpoints moved are satisfied by this first load, arcs entering a moved item are similarly satisfied there, and moving only the predecessor of an arc earlier cannot violate it. Remove trailing empty bins, if any, but retain intermediate empty bins. The resulting sequence $\widetilde{\boldsymbol{\mathcal{L}}}\in\mathscr{C}(S;\widehat{\mathcal{L}})$ is no longer than $\boldsymbol{\mathcal{L}}$. Because the same fixed $\widehat{\mathcal{L}}$ applies to every original sequence, it dominates $\mathcal{L}$, as claimed. For repeated application of the rule, compare loads lexicographically by $W(\mathcal{L})$, then $\sum_{u\in\mathcal{L}}\sum_{s\in\mathcal{I}}(\theta_{us}+1)$, and finally their $0/1$ item-incidence vectors in increasing item-index order. Every permitted replacement strictly increases this order, using the specified index tiebreaker when weights and labels coincide; a strict extension increases the total weight. Since the set of feasible maximal next loads is finite and the order strictly increases, following the replacements from any discarded load eventually reaches a load not discarded by this rule. By transitivity of dominance, this final load dominates the original one, establishing that the corresponding branch can be discarded without losing an optimal solution. \hfill\Halmos

\section{Experimental Settings}\label{app:experimental-settings}

\subsection{Computational Environments and Result Sources}\label{app:baseline-platforms}

The experiments with \solver\ were conducted using macOS 15.7.4 with compiler options \texttt{-O3 -DNDEBUG -mcpu=native -flto}. The Apple M4 Pro used in Section~\ref{sec:experiments} has 14 cores. All processor ratings in this supplement were obtained from \href{https://www.cpubenchmark.net/singleThread.html}{PassMark Software}.

For the published SALBP-I comparisons, the BBR12 results are in Table 1 of \citet{sewell2012branchappendix}. These were obtained using one core of an Intel Core 2 Duo T7200 processor with a clock speed of 2.0 GHz, 3.25 GB of memory, and a PassMark rating of 744. The numbers of instances proved optimal and the time limits for SALOME97 are taken from Table 2 of \citet{morrison2014application} and the official companion results of \citet{otto2013systematic}. These sources do not identify the original processor, memory, or thread count. BBR14 results come from \citet{morrison2014application} and its official supplement. These runs used one core of an Intel Core i7-930 processor at 2.8 GHz and 12 GB of memory. Average CPU times for Otto-20, the permuted Otto-50 set, and Otto-1000 exclude the time needed for search-tree memory initialization.

The results for I-SALOME20, I-BDP20, I-BBR20$'$, and I-BBR20 come from Table 16 in \citet{li2020comparative}, which also provides the BBR14 RPD for Otto-1000. These four improved methods use a single virtual processor and 8 GB of memory on a server equipped with Intel Xeon E5-2680 v2 processors running at 2.8 GHz. The host CPU rating does not measure the performance allocated to the virtual processor.

For Scholl BPP-P, CPLEX 12 and BB12 results come from Table 5 of \citet{dell2012binappendix}; CPLEX solves the compact model in that paper after first-fit initialization. Both use an Intel Pentium 3 GHz processor with 2 GB of memory, but the exact processor model and CPLEX thread count are unspecified. The \enumsixteen\ results are from Tables 1--4 of \citet{pereira2016proceduresappendix}, using one core of an unspecified eight-core Intel Xeon 2.66 GHz processor with 32 GB of memory. Neither description permits a reliable single-thread CPU rating.

The CPLEX results for the Otto sets across all three problem classes and the BM-ILS results for both BPP-GP precedence-weight sets are taken from Table 5 of \citet{kramer2017batchingappendix}. They were obtained using a single thread of an Intel Xeon E5530 2.4 GHz processor with 24 GB of memory. CPLEX 12.4 was initialized using first-fit. The \enumseventeen\ Otto-100 results come from Table 4 of \citet{kramer2017batchingappendix}, which reports new runs supplied by Pereira after implementation issues were fixed. These runs use a single thread on the i7-930 platform described above.

The BCP22 results for Otto SALBP-I, BPP-P, and BPP-GP with weights in $\{0,1\}$ and $n\in\{20,50,100\}$ come from Table 8 of \citet{letelier2022timelagsappendix}. Their C++ implementation uses BaPCod and CPLEX 12.8 on an Intel Xeon E5-2680 v3 processor running at 2.50 GHz with 128 GB of memory. Each run uses one thread.

\subsection{Time-Limit Settings}\label{app:time-limits}

Time limits were fixed before the reported experiments. For each benchmark group in the comparisons with published results, let $\mathcal{S}_{\mathrm{cmp}}$ denote the selected reference set against which the fixed time limit is assessed, rather than the full set of compared methods. For each $a\in\mathcal{S}_{\mathrm{cmp}}$, let $T_a$ and $R_a$ denote its reported time limit and identifiable single-thread CPU rating, respectively. Let $T_{\mathrm P}$ denote the per-instance time limit for \solver, and let $R_{\mathrm P}=4548$ denote its CPU rating.
The main targets of the 350- and 1000-second comparisons are competitive dedicated exact methods evaluated under longer time limits. CPLEX remains a published baseline but is excluded from these reference sets because its shorter limit would restrict those comparisons despite its weaker reported performance on the harder instances.

For Scholl BPP-P, the reported processor descriptions do not permit reliable single-thread CPU ratings, so $\mathcal{S}_{\mathrm{cmp}}=\varnothing$ and a fixed 1000-second limit is used. For all other groups, the reference set is nonempty and the selected limit satisfies
\begin{equation}
T_{\mathrm P}\le\frac{\min_{a\in\mathcal{S}_{\mathrm{cmp}}}\{T_aR_a\}}{R_{\mathrm P}}.
\label{eq:supp-cpu-budget}
\end{equation}

Table~\ref{tab:reference-time-calculations} gives the time limits, CPU ratings, and calculations for the reference methods. Table~\ref{tab:time-limit-settings} specifies each benchmark group's reference set, the resulting upper limit from Equation~\eqref{eq:supp-cpu-budget}, and the selected time limit.

\begin{table}[!htbp]
\centering
\caption{Time-Limit Calculations for the Reference Methods}
\label{tab:reference-time-calculations}
\fontsize{9pt}{11pt}\selectfont
\renewcommand{\arraystretch}{1.1}
\setlength{\tabcolsep}{3pt}
\begin{tabular*}{\textwidth}{@{\extracolsep{\fill}}lrrrr@{}}
\toprule
Method & $T_a$ (s) & $R_a$ & $T_aR_a$ & $T_aR_a/R_{\mathrm P}$ (s) \\
\midrule
CPLEX (Otto) & 300 & 1152 & 345,600 & 75.99 \\
BBR14 & 3600 & 1271 & 4,575,600 & 1006.07 \\
\enumseventeen & 3600 & 1271 & 4,575,600 & 1006.07 \\
I-SALOME20 & 900 & 1783 & 1,604,700 & 352.84 \\
I-BDP20 & 900 & 1783 & 1,604,700 & 352.84 \\
I-BBR20$'$ & 900 & 1783 & 1,604,700 & 352.84 \\
I-BBR20 & 900 & 1783 & 1,604,700 & 352.84 \\
BCP22 & 3600 & 1794 & 6,458,400 & 1420.05 \\
\bottomrule
\end{tabular*}
\par\vspace{3pt}
\begin{minipage}{\textwidth}
\fontsize{8pt}{11pt}\selectfont
\textit{Notes.} Only methods appearing in the selected reference sets are listed; $R_{\mathrm P}=4548$. Values in the last column are rounded to two decimal places.
\end{minipage}
\end{table}

\begin{table}[!htbp]
\centering
\caption{Reference Methods and Time Limits}
\label{tab:time-limit-settings}
\fontsize{9pt}{11pt}\selectfont
\renewcommand{\arraystretch}{1.1}
\setlength{\tabcolsep}{3pt}
\begin{tabular*}{\textwidth}{@{\extracolsep{\fill}}llllrr@{}}
\toprule
Problem & Benchmark set & $n$ & \shortstack[l]{Reference methods\\$\mathcal{S}_{\mathrm{cmp}}$} & \shortstack[r]{Upper\\limit/s} & \shortstack[r]{Selected\\limit/s} \\
\midrule
\multirow[c]{7}{*}{SALBP-I} & Scholl & All & I-BBR20 & 352.84 & 350 \\
 & Otto & 20 & BBR14, BCP22 & 1006.07 & 1000 \\
 & Otto base & 50 & BCP22 & 1420.05 & 1000 \\
 & Otto permuted & 50 & BBR14 & 1006.07 & 1000 \\
 & Otto & 100 & Improved methods, BCP22 & 352.84 & 350 \\
 & Otto & 250, 500, 750 & CPLEX & 75.99 & 75 \\
 & Otto & 1000 & Improved methods & 352.84 & 350 \\
\midrule
\multirow[c]{4}{*}{BPP-P} & Scholl & All & $\varnothing$ & -- & 1000 \\
 & Otto & 20, 50 & CPLEX, BCP22 & 75.99 & 75 \\
 & Otto & 100 & \enumseventeen, BCP22 & 1006.07 & 1000 \\
 & Otto & 250, 500, 750, 1000 & CPLEX & 75.99 & 75 \\
\midrule
\multirow[c]{3}{*}{BPP-GP} & Otto, weights $\{0,1\}$ & 20, 50, 100 & CPLEX, BCP22 & 75.99 & 75 \\
 & Otto, weights $\{0,1\}$ & 250, 500, 750, 1000 & CPLEX & 75.99 & 75 \\
 & Otto, weights $\{0,1,2,3\}$ & All & CPLEX & 75.99 & 75 \\
\bottomrule
\end{tabular*}
\par\vspace{3pt}
\begin{minipage}{\textwidth}
\fontsize{8pt}{11pt}\selectfont
\textit{Notes.} The improved methods are I-SALOME20, I-BDP20, I-BBR20$'$, and I-BBR20. The 525-instance Otto-50 SALBP-I base set inherits its 1000-second limit from the 5250-instance permuted set that contains it; this limit also satisfies Equation~\eqref{eq:supp-cpu-budget}.
\end{minipage}
\end{table}

Note that CPU ratings are used to assess the fixed time limits rather than normalize computing times. They do not establish equal computational effort across different architectures, virtualized environments, or algorithms.

\begingroup
\SingleSpacedXI
\putbib
\endgroup
\end{bibunit}

\end{APPENDICES}

\end{document}